%% file: ms.tex
\documentclass[preprint]{aastex}

\usepackage{amsmath,url,natbib}
\input{usersetcommands}

\shorttitle{A Possible Transiting T-Tauri Planet}
\shortauthors{van~Eyken et al.}

\begin{document}

\title{The PTF Orion Project: a Possible Planet Transiting a T-Tauri Star}

\author{Julian~C.~van Eyken,\altaffilmark{1,17,18}
  David~R.~Ciardi,\altaffilmark{1,18}
  Kaspar~von~Braun,\altaffilmark{1} Stephen~R.~Kane,\altaffilmark{1}
  Peter~Plavchan,\altaffilmark{1} 
  Chad~F.~Bender,\altaffilmark{2,3} Timothy~M.~Brown,\altaffilmark{4}
  Justin~R.~Crepp,\altaffilmark{5} Benjamin~J.~Fulton,\altaffilmark{4}
  Andrew~W.~Howard,\altaffilmark{6} Steve~B.~Howell,\altaffilmark{7,18}
  Suvrath~Mahadevan,\altaffilmark{2,3}
  Geoffrey~W.~Marcy,\altaffilmark{6} Avi~Shporer,\altaffilmark{4}
  Paula~Szkody,\altaffilmark{8,18}
  Rachel~L.~Akeson,\altaffilmark{1}
  Charles~A.~Beichman,\altaffilmark{1}
  Andrew~F.~Boden,\altaffilmark{9} Dawn~M.~Gelino,\altaffilmark{1}
  D.~W.~Hoard,\altaffilmark{10} Solange~V.~Ram\'{i}rez,\altaffilmark{1}
  Luisa~M.~Rebull,\altaffilmark{10} John~R.~Stauffer,\altaffilmark{10}
  Joshua~S.~Bloom,\altaffilmark{6} S.~Bradley~Cenko,\altaffilmark{6}
  Mansi~M.~Kasliwal,\altaffilmark{5}
  Shrinivas~R.~Kulkarni,\altaffilmark{5}
  Nicholas~M.~Law,\altaffilmark{11} Peter~E.~Nugent,\altaffilmark{12}
  Eran~O.~Ofek,\altaffilmark{13,19}
  Dovi~Poznanski,\altaffilmark{14}
  Robert~M.~Quimby,\altaffilmark{15} Richard~Walters,\altaffilmark{5}
  Carl~J.~Grillmair,\altaffilmark{10} Russ~Laher,\altaffilmark{10}
  David~B.~Levitan,\altaffilmark{16} Branimir~Sesar,\altaffilmark{5} \and
  Jason~A.~Surace\altaffilmark{10}}

\email{vaneyken@ipac.caltech.edu}

\altaffiltext{1}{NASA Exoplanet Science Institute, California Institute of
  Technology, 770 South Wilson Avenue, M/S~100-22, Pasadena, CA,
  91125, USA}

\altaffiltext{2}{Department of Astronomy \& Astrophysics, The
  Pennsylvania State University, University Park, PA 16802, USA}

\altaffiltext{3}{Center for Exoplanets and Habitable Worlds, The Pennsylvania State
  University, University Park, PA 16802, USA}

\altaffiltext{4}{Las Cumbres Observatory Global Telescope, Goleta, CA 93117, USA}

\altaffiltext{5}{Cahill Center for Astrophysics, California Institute of
Technology, Pasadena, CA, 91125, USA}

\altaffiltext{6}{Department of Astronomy, University of California, Berkeley, CA 94720-3411, USA}

\altaffiltext{7}{NASA Ames
  Research Center, M/S 244-30, Moffett Field, CA 94035, USA}

\altaffiltext{8}{Department of Astronomy, University of
   Washington, Box 351580, Seattle, WA 98195, USA}

 \altaffiltext{9}{Caltech Optical Observatories, California Institute
   of Technology, Pasadena, CA 91125, USA}

 \altaffiltext{10}{Spitzer Science Center, M/S 220-6, California
   Institute of Technology, Jet Propulsion Laboratory, Pasadena,
   CA 91125, USA}

 \altaffiltext{11}{Dunlap Institute for Astronomy and Astrophysics,
   University of Toronto, 50 St. George Street, Toronto M5S 3H4,
   Ontario, Canada}

 \altaffiltext{12}{Computational Cosmology Center, Lawrence Berkeley
   National Laboratory, 1 Cyclotron Road, Berkeley, CA 94720, USA}

\altaffiltext{13}{Benoziyo Center for Astrophysics, Weizmann Institute of Science,
76100 Rehovot, Israel}

\altaffiltext{14}{School of Physics and Astronomy, Tel Aviv University, Tel Aviv 69978, Israel}

\altaffiltext{15}{IPMU, University of Tokyo, Kashiwanoha 5-1-5, Kashiwa-shi, Chiba, Japan}

 \altaffiltext{16}{Division of Physics, Mathematics and Astronomy, California Institute of
   Technology, Pasadena, CA 91125, USA}

\altaffiltext{17}{Currently at Department of Physics, UC Santa Barbara, Santa
Barbara CA 93106; vaneyken@physics.ucsb.edu}

 \altaffiltext{18}{Visiting Astronomer, Kitt Peak National
   Observatory, National Optical Astronomy Observatory, which is
   operated by the Association of Universities for Research in
   Astronomy, Inc. (AURA) under cooperative agreement with the
   National Science Foundation}

\altaffiltext{19}{Einstein Fellow}

\begin{abstract}
\input{abstract}
\end{abstract}

\keywords{Open clusters and associations: individual (25 Ori) -- planets and satellites:
  detection -- stars: individual (2MASS J05250755+0134243 /
  CVSO 30 / PTFO 8-8695 /  PTF1 J052507.55+013424.3)  -- stars: pre-main sequence}

%------------------------------------------------------------------
\input{introduction}

\input{observations}
\input{discussion}

\input{conclusions}
%------------------------------------------------------------------
\acknowledgments
\input{acknowledgments}
%------------------------------------------------------------------

%------------------------------------------------------------------

% Make List of References (BibTeX implemented)
\bibliographystyle{apj} 
\bibliography{apj-jour,ptfbibliography}

\clearpage

\end{document}

%% file: usersetcommands.tex
\newcommand{\kms}{\,{\rm km}\,{\rm s}^{-1}}
\newcommand{\ms}{\,{\rm m\,s^{-1}}}
\newcommand{\pc}{\,{\rm pc}}

\newcommand{\Msun}{\, M_{\odot}}
\newcommand{\Lsun}{\,{L_{\odot}}}
\newcommand{\Rsun}{\,{R_{\odot}}}
\newcommand{\Rjup}{\,{R_\mathrm{Jup}}}

\newcommand{\ang}{\,\text{\AA}}

\newcommand{\m}{\,{\rm m}}

\newcommand{\Mjup}{\, M_{\rm Jup}}

\newcommand{\days}{\,{\rm d}}
\newcommand{\Myr}{\,{\rm Myr}}

\newcommand{\AU}{\,{\rm AU}}

\newcommand{\vmag}{\,{\rm mag}}
\newcommand{\mmag}{\,{\rm mmag}}

\newcommand{\K}{\,{\rm K}}

\newcommand{\pix}{\,{\rm pix}}

\newcommand{\mas}{\,{\rm mas}}
\newcommand{\rchisq}{\chi^2_{\rm red}}
\newcommand{\inch}{\arcsec}
\newcommand{\seconds}{\,\mathrm{s}}

%% file: abstract.tex
We report observations of a possible young transiting planet orbiting
a previously known weak-lined T-Tauri star in the 7--$10\Myr$-old
Orion-OB1a/25-Ori region. The candidate was found as part of the
Palomar Transient Factory (PTF) Orion project. It has a photometric
transit period of $0.448413 \pm 0.000040$ days, and appears in both
2009 and 2010 PTF data. Follow-up low-precision radial velocity
observations and adaptive-optics imaging suggest that the star is not
an eclipsing binary, and that it is unlikely that a background source
is blended with the target and mimicking the observed transit.
Radial-velocity observations with the Hobby-Eberly and Keck telescopes
yield a radial velocity that has the same period as the photometric
event, but is offset in phase from the transit center by $\approx
-0.22$ periods. The amplitude (half range) of the radial velocity
variations is $2.4\kms$ and is comparable with the expected radial
velocity amplitude that stellar spots could induce. The radial
velocity curve is likely dominated by stellar spot modulation and
provides an upper limit to the projected companion mass of
$M_\mathrm{p}\sin i_{\mathrm{orb}} \lesssim 4.8\pm 1.2\Mjup$; when combined
with the orbital inclination, $i_{\mathrm{orb}}$, of the candidate
planet from modeling of the transit lightcurve, we find an upper limit
on the mass of the planetary candidate of $M_\mathrm{p} \lesssim
5.5\pm 1.4\Mjup$. This limit implies that the planet is orbiting close
to, if not inside, its Roche limiting orbital radius, so that it may
be undergoing active mass loss and evaporating.

%% file: introduction.tex
\section{Introduction}\label{sec:intro}

The Palomar Transient Factory (PTF) Orion project is a study
within the broader PTF survey aimed at searching for photometric
variability in the young Orion region, with the primary goal of
finding young extrasolar planets \citep{PTFOrionOverview}. The project
is based on a set of intensive high-cadence ($\approx
70$--$90\seconds$) observations of a single $7.26\,\mathrm{deg}^2$
field centered on the known \object{25 Ori} association, which lies in
the Orion OB1a region and has an estimated age of 7--10$\Myr$
\citep{Briceno2005,Briceno2007}.

Typical young circumstellar disk lifetimes are on the order of
5--$10\Myr$ \citep{Hillenbrand2008}, and it is during this time that
the bulk of the formation and migration of planets is expected to
occur. The youngest exoplanets have been found via direct detection
(\citealp[e.g., LkCa 15,][]{Kraus2012}; \citealp[the free-floating
planetary-mass object in the $\rho$ Oph cloud,][]{Marsh2010});
however, little is known empirically about exoplanets during the first
few millions of years of their lives, and the goal is to fill the
observational gap and begin to provide constraints on theories of
planet formation and evolution \citep[see,
e.g.,][]{ArmitagePlanetFormation}. The stars in the 25 Ori/OB1a region
should be at or just beyond the point of disk dissipation, consistent
with the high ratio of weak-lined to classical T-Tauri stars found
there \citep{Briceno2007}. We can therefore look for planet transits
at the time when they may first become observable without their
signatures being swamped by the extreme variability characteristic of
the younger classical T-Tauri stars.  In so doing, we can begin to
investigate the frequency of planets at these ages; the timescales for
their evolution; the timescales of their migration with respect to the
star's evolution; and, through measurements of the transit depths,
probe empirically their mean densities and the extent of their
atmospheres, which are expected to be inflated at these early stages
\citep[][and references therein]{Fortney2010}.

Several surveys have been undertaken to search for close-in young
exoplanets, though many are radial-velocity searches
\citep[e.g.,][]{Esposito2006, Paulson2006, Setiawan2007, TWHya,
  Huerta2008, Crockett2011, Nguyen2012}. A transit search
\citep[e.g.,][]{Monitor,Miller2008,YETI} has the advantage of being
able to search many more stars simultaneously, down to fainter
magnitudes and with a faster cadence. Combining transit photometry
with spectroscopic information can provide a wealth of information
that is unavailable with non-transiting planets, from masses and radii
to constraints on atmospheric composition.

An outline of the PTF Orion project is given by
\citet{PTFOrionOverview}, along with some of the first results
concerning binary stars and T-Tauri stars. The broader PTF survey is
described in detail by \citet{PTFtechnical} and \citet{PTFsurvey},
with a summary of the first year's performance by
\citet{Law2010}. Here we report a young planet candidate found
orbiting a known M3 pre-main-sequence (PMS) weak-lined T-Tauri star
(TTS) within the PTF Orion field. Our observations are outlined in
Section \ref{sec:observations}; Section \ref{sec:results} discusses
the photometric and spectroscopic results and some implications. The
main conclusions are summarized in section \ref{sec:conclusions}.

%% file: observations.tex
\section{Observations}\label{sec:observations}

\subsection{PTF Photometry}\label{sec:ptfphotometry}

\begin{figure}[tbp]
%\centering
\epsscale{0.4}
\plotone{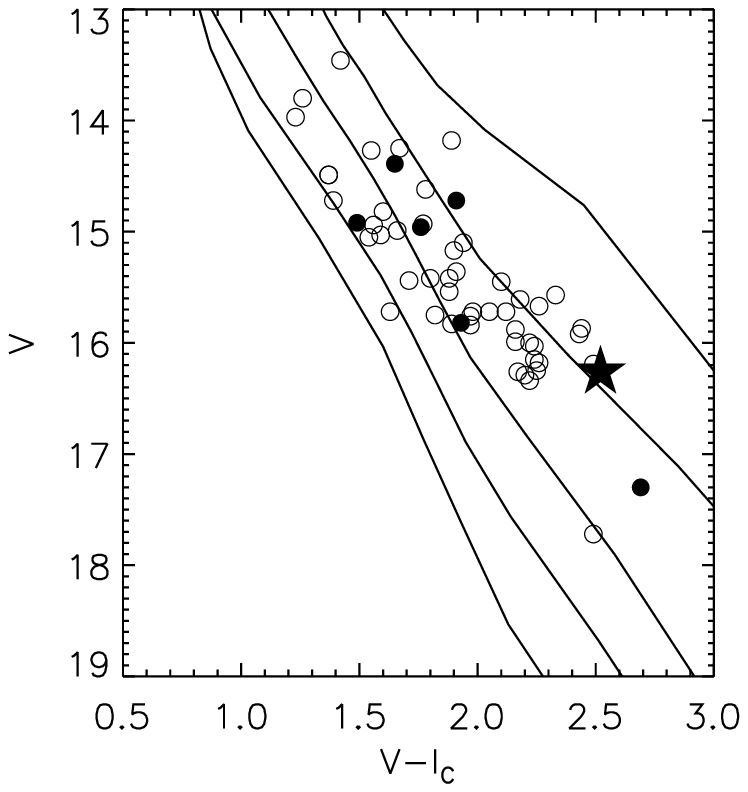}
\caption{\label{fig:CMD}Color-magnitude diagram following
  \citet[Figure 7,][]{Briceno2005}, highlighting PTFO 8-8695 (CVSO 30)
  -- the star symbol -- in relation to the other T-Tauri stars
  discovered therein (see section \ref{sec:ptfphotometry}). PTFO
  8-8695 lies at the younger end of the distribution. Open
  circles indicate WTTS; filled circles indicate CTTS. Solid lines
  indicate, from the top, 1, 3, 10, and $30\Myr$ isochrones, and the
  zero-age main sequence, according to the models of
  \citet{SiessModels} at a distance of $330\pc$. Photometric
  measurements are reproduced from \citet{Briceno2005}. Reddening and extinction
  are neglected.}
\end{figure}

PTF Orion data were obtained using the Palomar $48\inch$ Samuel Oschin
telescope during the majority of the clear nights between 2009
December 1 and 2010 January 15, whenever the field was above an
airmass of 2.0. All observations were in the $R$ band, and of the 40
nights dedicated, 14 yielded usable data, the remainder being lost
primarily to poor weather. Light curves were obtained for $\sim
110,000$ sources within the field, the top $\sim 500$ most variable of
which were inspected visually. The observations and the PTF Orion
differential photometry pipeline and data reduction details are
described fully in \citet{PTFOrionOverview}. Specifics of the more
general PTF data reduction pipeline are described by R.~Laher et
al. (2012, in preparation) and \citet{Ofek2012}. Among the inspected
sources, one, \object{PTFO 8-8695} (\object{2MASS J05250755+0134243}),
showed periodic shallow transit-like events with a shape and depth
suggestive of a planetary companion, superposed on a larger-scale
quasi-periodic variable light curve. The source has previously been
identified as a weak-line T-Tauri star (WTTS) associated with the
Orion OB1a region by \citet[\object{CVSO 30},][]{Briceno2005}, and
\citet{Hernandez2007}. A summary of the main previously determined
properties of the primary star is given in Table
\ref{tbl:starproperties}. In Figure \ref{fig:CMD} we show a
color-magnitude diagram indicating PTFO 8-8695 in relation to the
other T-Tauri stars in the OB1a region discovered by
\citet{Briceno2005}. Although classified as a WTTS, it lies at the
younger end of the distribution. This is consistent with its its
relatively strong $\mathrm{H}_\alpha$ emission, which in fact places
it on the borderline with classical T-Tauri stars (CTTS) according to
the classification scheme of \citet{Briceno2005}.

\input{table-starproperties}

The same field was again observed in the same way over seven clear
nights between 2010 December 8--17. The transit events were again
evident with the same period and depth.
Owing to improvements in the PTF image processing software, better
weather, and mitigation of the CCD fogging effect seen in the previous
year's data \citep[see][]{PTFOrionOverview,Ofek2012}, the RMS noise
floor for the brightest stars in the 2010 December data set improved from
$\approx 4$ to $\approx 3\mmag$, with less evidence of systematic
effects.\footnote{Independent differential photometric analyses of
  other PTF data, and respective precision levels, are discussed in
  \citet{Agueros2011} and \citet{Levitan2011}.}
Figures 2 and 3 show the light curves obtained in the first and second
years of observations with PTF. The quasi-periodic stellar variability is
evident, along with sporadic low-level flaring, consistent with the
star's young age and late spectral type. Visual inspection of the
curves revealed the additional regular periodic transit-like signature,
with a depth of $\approx 3$--4\%, superposed on the stellar
variability. We initially determined an approximate period ($\approx 0.45$ days)
and transit duration ($\approx 2$ hours) by hand, in order to locate all
the transit windows.  A detailed discussion of the light curves and
derived properties is given in Section \ref{sec:photometry}. The
regularity and planet-like characteristics of the transit events led
us to obtain follow-up observations.

\begin{figure*}[tbp]
%\centering
\epsscale{0.9}
\plotone{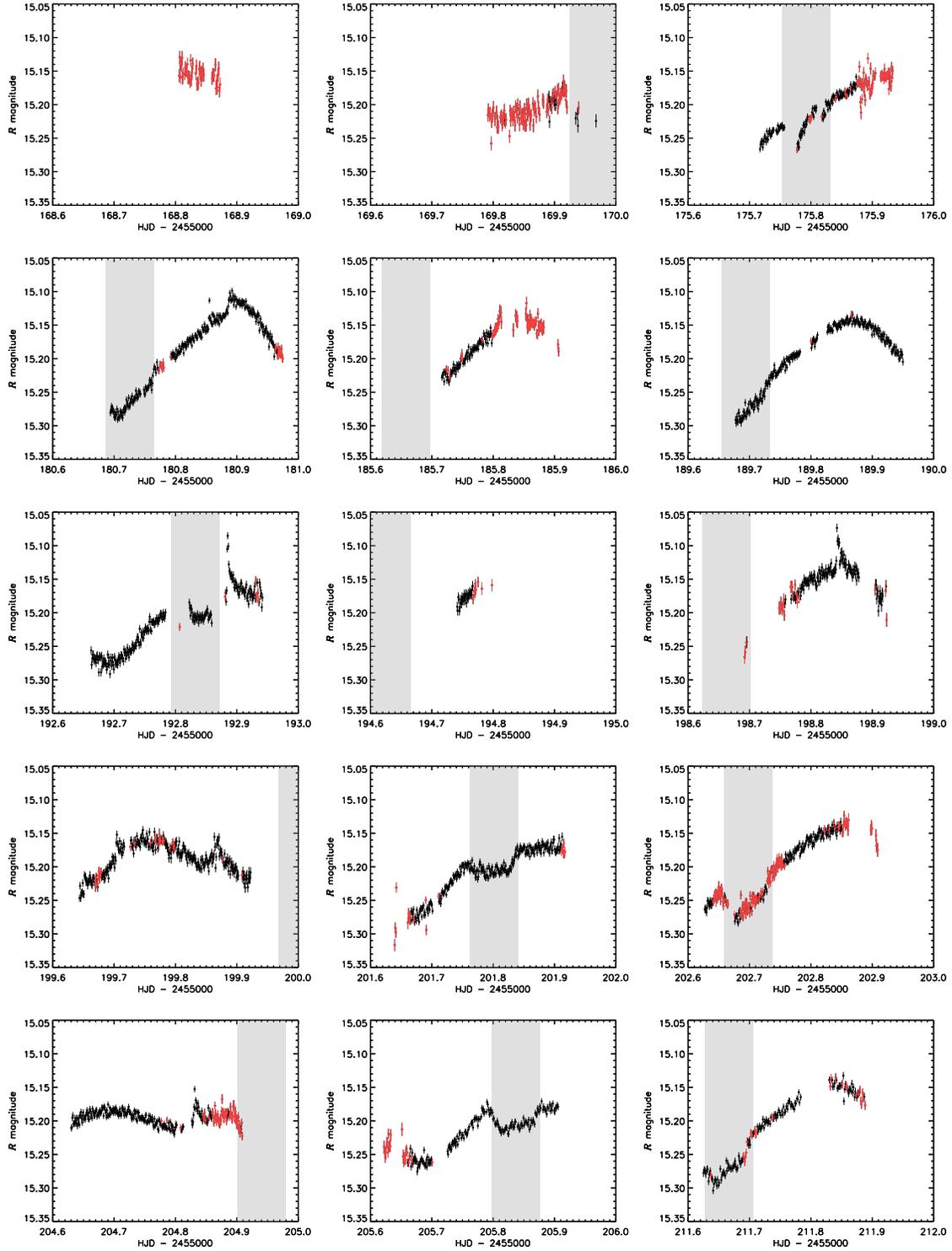}
\caption{\label{fig:LC1} Photometric light curve for 2009 December
  1--2010 January 15 (Sections \ref{sec:ptfphotometry} and
  \ref{sec:LCandPeriods}). Gray regions indicate the transit windows,
  fixed at the measured transit period, width, and epoch of
  center-transit ($T_0$). Red points indicate data automatically
  flagged by the data reduction software as potentially non-optimal
  for various possible reasons (e.g., imperfect weather, evidence of
  contamination within the photometric aperture, etc.)}
\end{figure*}

\begin{figure*}[tbp]
%\centering
\epsscale{0.9}
\plotone{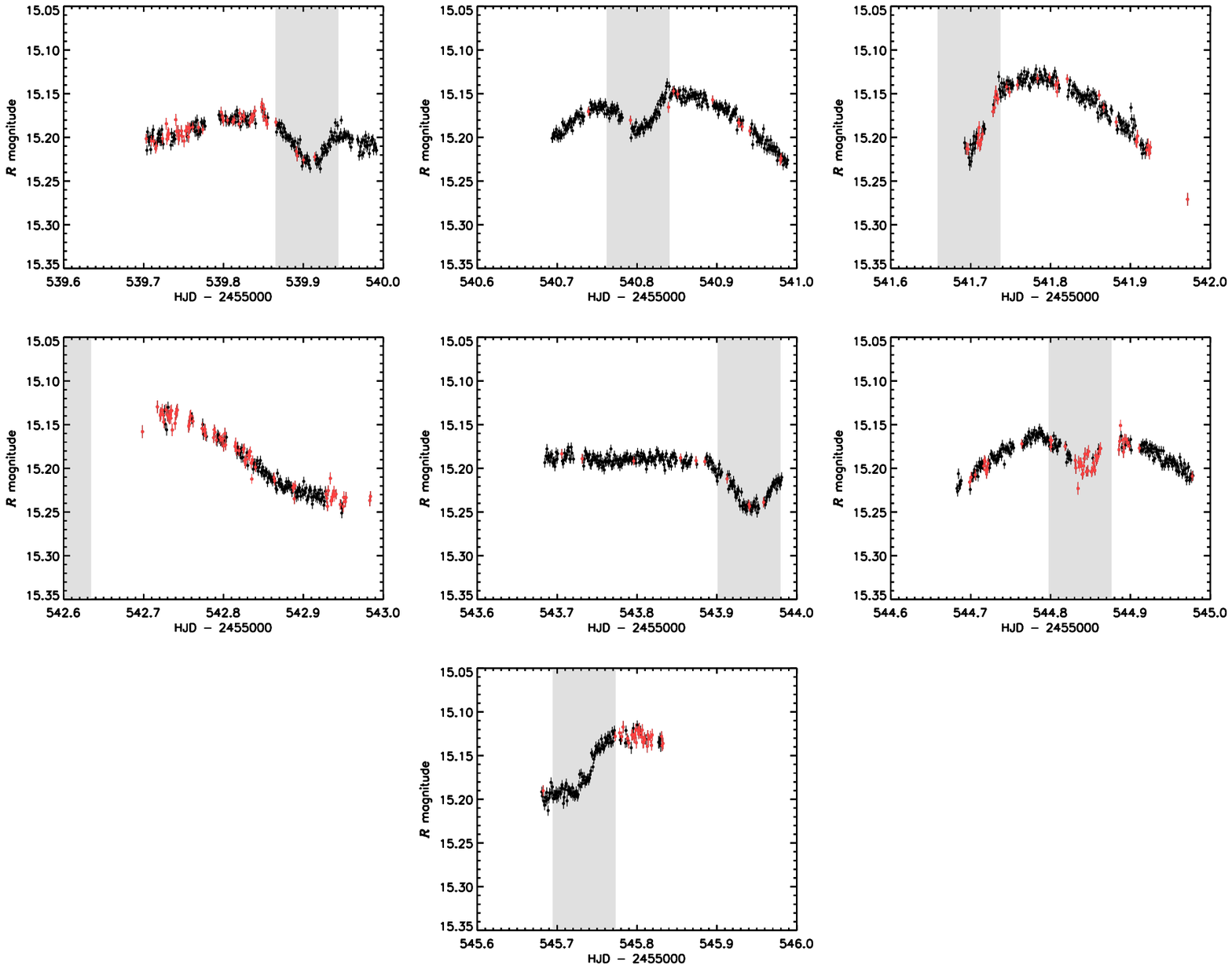}
\caption{\label{fig:LC2}As for Figure \ref{fig:LC1}, for 2010 December
  8--17.}
\end{figure*}

\subsection{Keck Adaptive Optic Imaging}\label{sec:AO}

The PTF imaging instrument is seeing limited and has a point-spread
function (PSF) with typical full-width-half-maximum (FWHM) of
$2.0\arcsec$, or $660\AU$ at the estimated distance of the Orion OB1a
and 25~Ori associations
\citep[$330\pc$,][]{Briceno2005,Briceno2007}. We obtained adaptive
optics (AO) images using the NIRC2 camera (PI K.~Matthews) on the
$10\m$ Keck II telescope in order to probe regions in the immediate
vicinity of the star and to search for any sources (false positives)
that could mimic the signal of a transiting planet, such as a nearby
eclipsing binary. At $R=15.2\vmag$, PTFO 8-8695 is sufficiently bright
for natural guide star observations and does not require the use of a
laser for atmospheric compensation. We locked the AO system control
loops onto the target using a frame rate of 41 Hz. The airmass was
1.29. We used the NIRC2 narrow camera setting to provide fine spatial
sampling ($10\mas \pix^{-1}$). Our observations consisted of 12
dithered \textit{H}-band images (3 coadds per frame, $10\,\mathrm{s}$ per
coadd), totaling 6 minutes of on-source integration time. Raw frames
were processed by cleaning hot pixels, subtracting background noise
from the sky and detector, and aligning and coadding the results.

\begin{figure}[tbp]
%\centering
%\epsscale{1.1}
\plotone{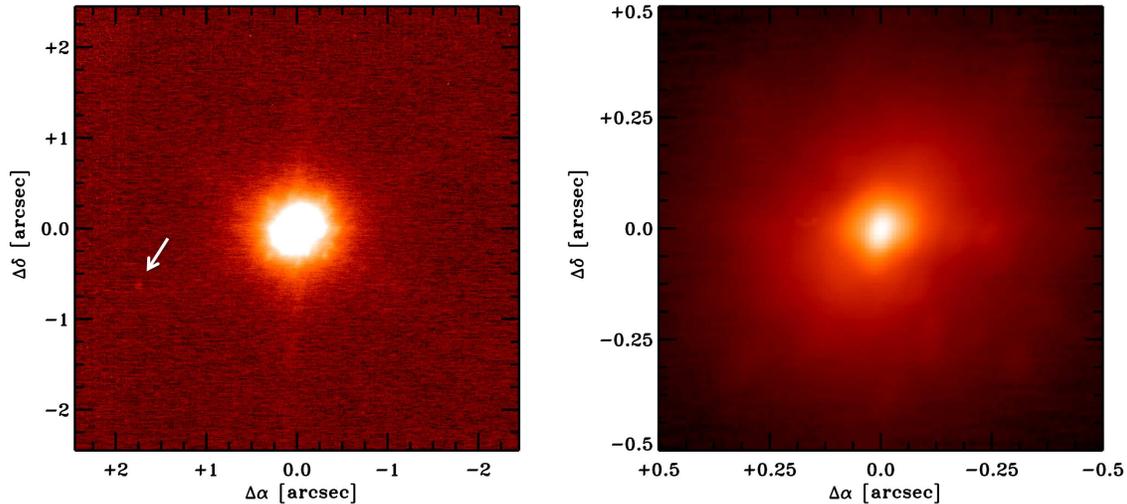}
\caption{\label{fig:AO} Adaptive optics imaging of PTFO 8-8695,
  showing the full image field (left), and detail around the central
  star (right), at two different logarithmic brightness stretches. A
  very faint source is detected to the south east of the central star
  at a separation of 1.8\arcsec, $6.96\vmag$ fainter than the central
  source, and unlikely to be capable of mimicking the observed
  transits (indicated by the arrow, left panel); otherwise no contaminants are
  detected. (Section \ref{sec:AO}.)}
\end{figure}

Figure \ref{fig:AO} shows the final reduced image using an effective
field-of-view of $4.9\arcsec \times 4.9\arcsec$, which corresponds to
$\approx 4.9$ PTF pixels on a side. The image shows no contaminants,
except for one faint source to the south east, at a separation of
$1.8\arcsec$ ($590\AU$ in the plane of the sky at the distance of
Orion OB1a/25~Ori). Our diffraction-limited images (FWHM $\approx
80\mas$) rule out additional off-axis sources down to a level of
$\Delta H=4.3$, 6.4, 8.9, and $9.1\vmag$ ($3\sigma$) at angular
separations of 0.25\arcsec, 0.5\arcsec, 1.0\arcsec, and $2.0\arcsec$
(83, 165, 330, and 660\pc) respectively. Assuming a color difference
of $0\vmag$, a faint, blended, 100\% eclipsing binary would have to be
within $\Delta H\approx 3.5$--$3.8\vmag$ of the primary star, or
brighter, to mimic the observed transit depth at $R$ of $\approx
3$--4\%. At $6.96\vmag$ (608 times) fainter than our target, the one
imaged contaminant is too faint to be a blended eclipsing
binary capable of mimicking the observed transits unless it is extremely blue
($R-H\lesssim -3.2$); in that event, however, it would be unlikely to be stellar in
origin in any case. Any other such binary would have to lie within
$0.25\arcsec$ of our target in order not to have been detected.

\subsection{LCOGT, KPNO 4\,m, and Palomar 200\arcsec~Vetting}\label{sec:LCOGT}

\begin{figure}[tbp]
\centering
\epsscale{1.01}
\plottwo{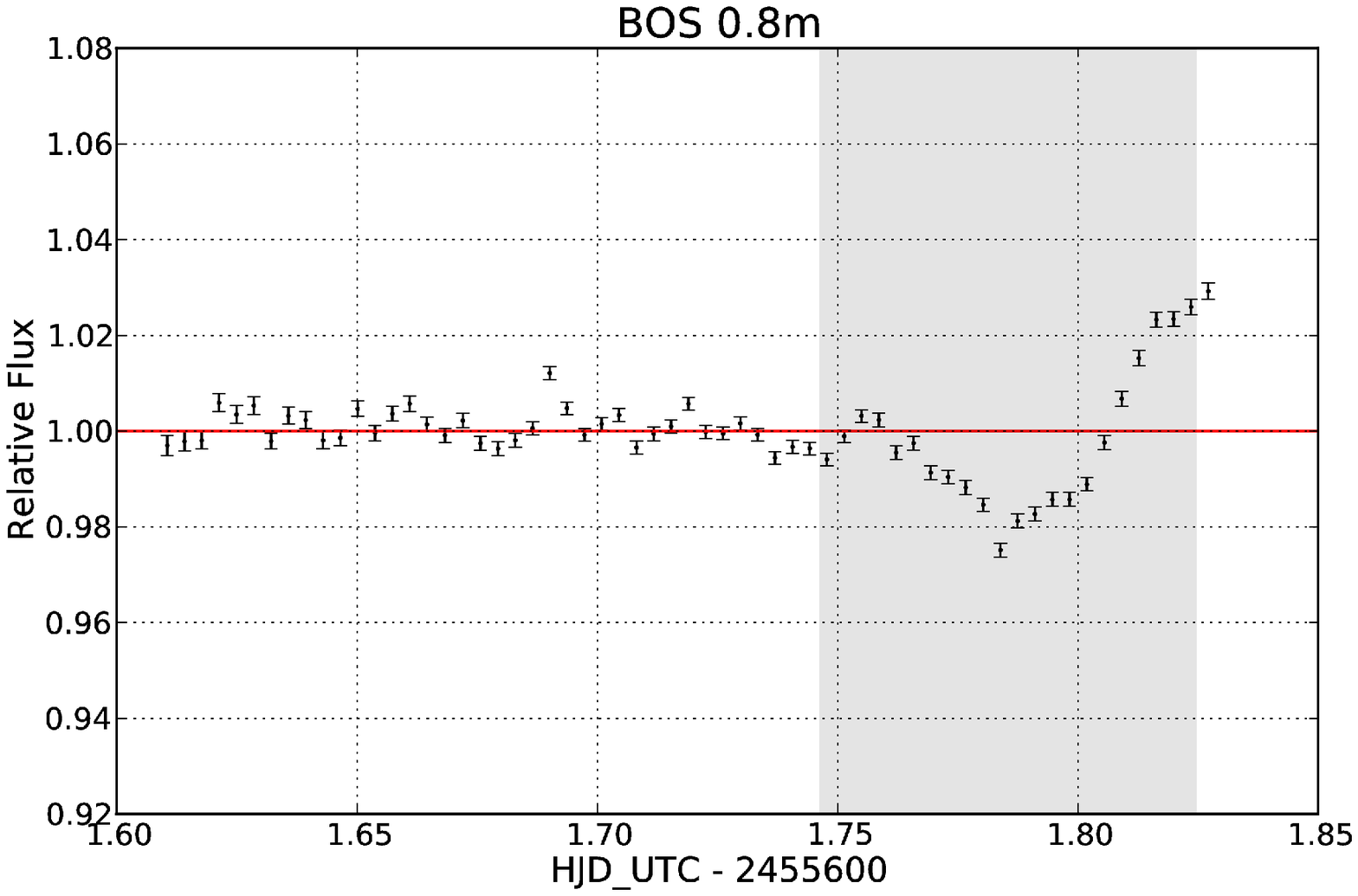}{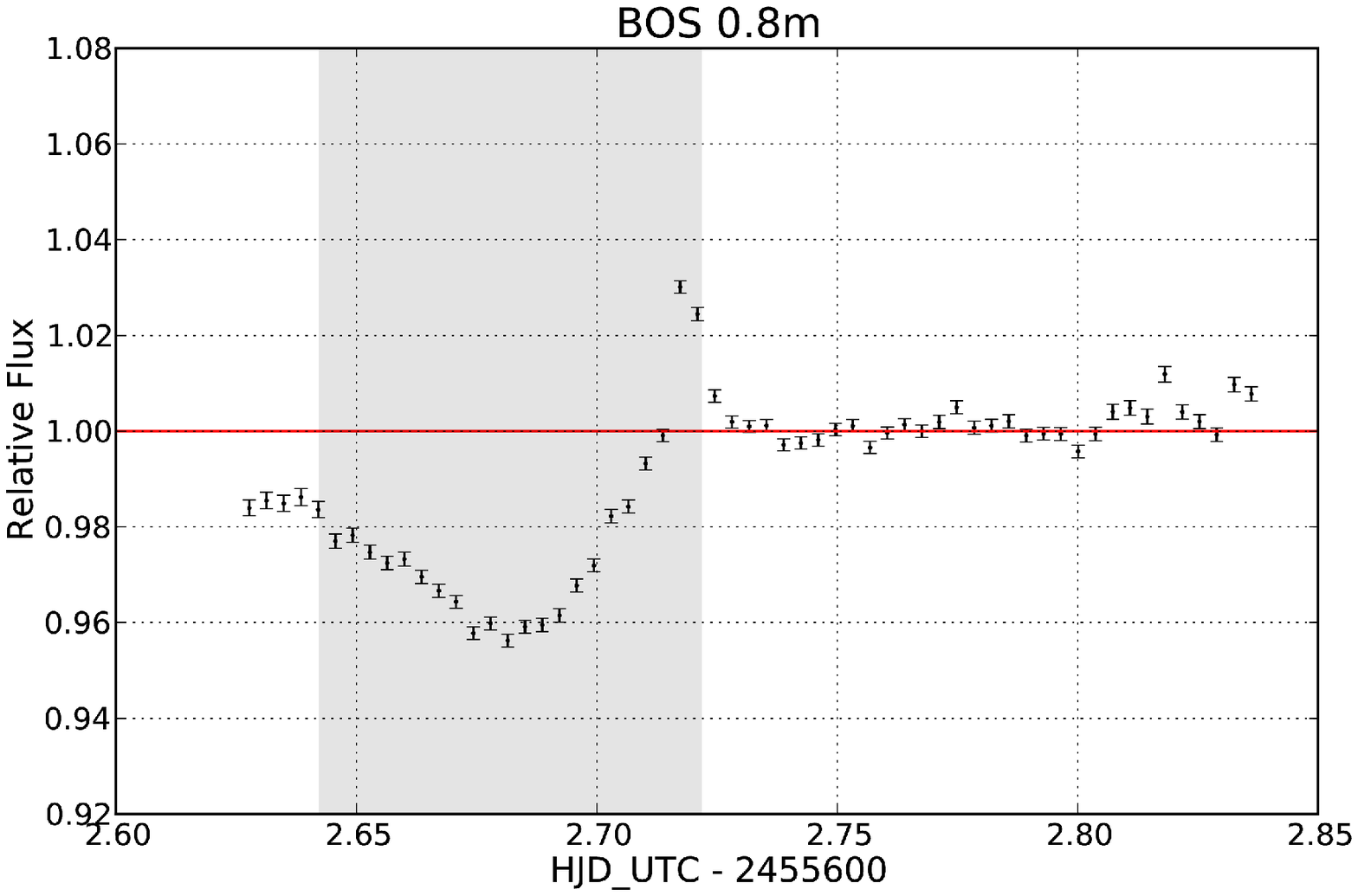}
\caption{\label{fig:LCOGT}Transit observations obtained with the LCOGT
  BOS 0.8\,m telescope (clear filter), on the nights of 2011
  February 8 and 9 (Section \ref{sec:LCOGT}). Gray regions again
  indicate the predicted transit windows using the period and $T_0$
  obtained with the PTF data. The second night shows a possible small
  flare at the end of the transit. Note that in addition to stellar
  variability, some of the systematic trends may also be air-mass
  related.}
\end{figure}

Photometric observations were obtained in 2011 February with the Las
Cumbres Observatory Global Telescope Network (LCOGT), using the $2\m$
Faulkes North and South telescopes, and the $0.8\m$ Byrne Observatory
telescope at Sedgwick Reserve (BOS). We were able to confirm complete
transit detections on two separate nights with the BOS telescope with
a clear filter (see Figure \ref{fig:LCOGT}).  Observations in SDSS
$g^\prime$ and $i^\prime$ with the three telescopes were inconclusive,
and largely hampered by poor observing conditions and sparse phase
coverage.  The LCOGT observations enabled us to confirm the transit
event and its period independently of the PTF data, and, in
combination with the PTF data, we were able to establish a more
accurate and up-to-date transit ephemeris.

In addition we obtained low-precision followup radial velocity (RV)
observations with the KPNO 4\,m Mayall telescope and the Palomar
$200\inch$ Hale telescope over 3 and 2 consecutive nights,
respectively, to allow us to rule out a stellar-mass companion as the
cause of the transits. The separate confirmation of the transit event
and the lack of RV signature at the level of $\sim 10\kms$ suggested
that we had indeed detected a sub-stellar object, and we pursued additional
observations.

\subsection{HET and Keck Spectroscopy}\label{sec:HETkeck}

Following the AO vetting and low-precision RV followup, we obtained
Doppler radial velocity spectroscopy with both the High Resolution
Spectrograph \citep[HRS;][]{Tull1998} on the $9.2\m$ Hobby Eberly
Telescope \citep[HET;][]{Ramsey1998}, and the High Resolution Echelle
Spectrometer \citep[HIRES;][]{HIRES} on the $10\m$ Keck I telescope.
The goal was to detect or place an upper limit on any signal due to
reflex stellar motion caused by the orbiting companion.

The queue-schedule operation mode of the HET \citep{Shetrone2007}
allowed us to obtain four spectroscopic radial velocity observations
very quickly after analysis of the low-precision RV and LCOGT
photometric vetting data. The observations were timed on the basis of
prior knowledge of the transit ephemeris to best constrain any orbital
signature.

Observations from the fiber-fed HRS spectrograph and calibrated using
ThAr exposures provide radial velocity precision of better than $50\ms$
over timescales of several weeks \citep{Bender2012}. Seeing-induced
PSF changes are minimal due to the image scrambling properties of
optical fibers \citep{Heacox1987} and the HRS temperature is kept stable to
$\sim 0.01\,\mathrm{C}$. We used the 15k resolution mode of HRS, with
the red cross-disperser setting (316g7940), giving a wavelength
coverage of $6114$--$9861\ang$. The use of this lower resolution
setting enhances the efficiency of HRS without significantly degrading
the information content in the stellar spectrum (since the target
absorption lines were already known to be rotationally
broadened). Each observation was $1200\seconds$ in duration, except on
UT 2011-02-21 which was $1600\seconds$, and ThAr calibration frames
were taken immediately after each to track instrument drift. Data for
the four nights were reduced in IDL with a custom optimal extraction
pipeline that performs bias subtraction, flat fielding, order tracing
and extraction, and wavelength calibration of the extracted
one-dimensional spectra. A simultaneous sky-fiber was available with
this spectrograph setting, but we chose to avoid the additional
complexity inherent in performing sky subtraction with a fiber-fed
spectrograph, and instead masked areas affected by sky lines so that
they were not used in calculating the RVs. The achieved
signal to noise in each spectrum was approximately 40--50 per
resolution element, peaking in the \textit{I}-band.  We identified the
strongest absorption lines in an M2-type high resolution Kurucz
stellar model\footnote{\url{http://kurucz.harvard.edu}}
\citep{Kurucz2005}, and used these to select lines for fitting in the
measured spectra.

For the Keck data, we observed PTFO 8-8695 using the standard
procedures \citep[e.g.,][]{Howard2009} of the California Planet Search
(CPS) for HIRES.  The five observations over 10 days were each 500--$600
\seconds$ in duration and achieved a signal to noise ratio of 10--20 per
pixel (S/N = 20--40 per resolution element) in the \textit{V}-band.  We used the
``C2'' decker ($0.86\arcsec$ wide slit) for a spectral resolution of
$\sim$60,000. The total wavelength coverage was $\approx
3650$--$7970\ang$.

Owing to the faintness of the target and the relatively relaxed
precision requirements, we employed in both cases thorium-argon (ThAr)
emission lamps as the fiducial reference. This avoids the
light-throughput penalty incurred by the higher precision common-path
iodine gas absorption cell, which is better suited to brighter
targets.

Data reduction and analysis were performed independently at separate
institutions for the two data sets, using independently written
software, but following the same general principles. Visual inspection
of the spectra readily revealed strong rotational line broadening
($v_\mathrm{r}\sin i_* = 80.6\pm 8.1\kms$ -- see section
\ref{sec:vsini}), to the extent that only the very strongest of the
photospheric lines were evident above the noise level. 
The low signal-to-noise ration (S/N) and high stellar rotational velocity
resulted in a very small number of lines from which to measure the
radial velocity.  Using a cross-correlation analysis on the HRS spectra
to measure these broad, low S/N lines yielded a velocity uncertainty of
$\sim10$ -- 14$\kms$.  Such an analysis is very sensitive to bad pixels or
faint sky lines, which were difficult to remove in the low S/N regime.
In addition, a cross correlation strongly benefits by combining the
information content of multiple lines with fixed relative spacings.  In
our data, viable lines tended to be spaced far apart: typically there
were only one or two lines per spectral order, which negated much of the
advantage of using a cross-correlation.  Consequently, we adopted an
alternative approach of fitting individual stellar lines with Gaussian
functions, solving each line independently for its center, width, and
depth. This fitting procedure was much less sensitive than
cross-correlation to bad pixels or to the precise spectral window being
analyzed.  It also allowed us to mask out telluric absorption and
strong, narrow night sky emission lines.

The faintness of the target and the small semi-major axis of the
putative companion's orbit meant that very high RV precision was neither
anticipated nor required. Simple Gaussian+linear-term profiles
provided an adequate fit (i.e., reduced $\chi^2\sim 1$) owing to the
low S/N and heavy broadening (and probable blending) of the lines,
where more sophisticated models would have added unnecessary and
unconstrained extra parameters. A weighted average of the shifts in
the fit centroids with respect to the rest wavelengths of the stellar
lines provided the required Doppler shifts. Since substantially
differing S/N in the various lines precluded a simple
standard-deviation--based estimate of the measurement errors, errors
were instead estimated by propagating the formal errors from the
Gaussian fits to the lines. In the case of the Keck data, corrections
were made for measured variation in the telluric lines to account for
residual uncalibrated instrument drift, but these corrections were
found to be negligible. A total of 14 absorption features spanning
$\sim 5300$--$7700\ang$ were fit in the Keck data, and 14 spanning
$\sim 6100$--$8700\ang$ in the HET data. The measured RVs are listed
in table \ref{tbl:RVs}.

\input{table-RVs}

%% file: table-starproperties.tex
\begin{deluxetable}{rl}
%\tabletypesize{\scriptsize}
\tablecaption{PTFO 8-8695 Stellar Properties\label{tbl:starproperties}}
%\tablewidth{0pt}
\tablehead{\colhead{Property} & \colhead{Value}}
\startdata
Alternative designations &  \object{CVSO 30} \\
                         &  \object{2MASS J05250755+0134243} \\
                         &  \object{PTF1 J052507.55+013424.3} \\
$\alpha$ (J2000) & 05\fh25\fm07\fs55\\
$\delta$ (J2000) & +01\fdg34\farcm24.3\farcs\\
$V$ & 16.26\vmag\tablenotemark{a}\\
2MASS $J$ & $12.232\pm0.028\vmag$\tablenotemark{b}\\
2MASS $H$ & $11.559\pm0.026\vmag$\tablenotemark{b}\\
2MASS $K_S$ & $11.357\pm0.021\vmag$\tablenotemark{b}\\
Median $R$ & $15.19\vmag$\tablenotemark{c}\\
$R$ range & $0.17\vmag$ (min to max)\tablenotemark{c} \\%From first year, which has lower min/higher max.
$\mathrm{H_\alpha}$ equiv. width &  -11.40\AA\tablenotemark{a}  \\
LiI equiv. width & 0.40 \AA\tablenotemark{a} \\
Sp. Type & M3 (PMS weak-lined T-Tauri)\tablenotemark{a}\\
$T_{\mathrm{eff}}$ & $3470\K\tablenotemark{a} $ \\
$A_V$ & $0.12\vmag$\tablenotemark{a} \\
Luminosity & $0.25\Lsun$\tablenotemark{a} \\
Radius & $1.39\Rsun$\tablenotemark{a,}\tablenotemark{d} \\
Mass (Baraffe/Siess) & $0.44\Msun/0.34\Msun$\tablenotemark{a,}\tablenotemark{e}\\
Age (Baraffe/Siess) & $2.63\Myr/2.68\Myr$\tablenotemark{a,}\tablenotemark{e}\\
Distance & $\sim 330\pc$ (mean dist. to OB1a/25 Ori assoc.)\tablenotemark{a,}\tablenotemark{f}
\enddata
\tablenotetext{a}{\citet{Briceno2005}.}
\tablenotetext{b}{\citet{2MASS}.}
\tablenotetext{c}{From PTF Orion data (this paper).}
\tablenotetext{d}{cf. smaller value implied by PTF Orion transit measurements -- see Table \ref{tbl:transitparams}.}
\tablenotetext{e}{Reported by \citet{Briceno2005} based on comparison
  with \citet{BaraffeModels} and \citet{SiessModels} stellar models.}
\tablenotetext{f}{\citet{Briceno2007}}
\end{deluxetable}

%% file: table-RVs.tex
\begin{deluxetable}{lr@{$\pm$}lc}
%\tabletypesize{\scriptsize}
\tablecaption{Differential Radial Velocity Measurements\label{tbl:RVs}}
%\tablewidth{0pt}
\tablehead{
  \colhead{HJD} & \multicolumn{2}{c}{RV} & \colhead{Telescope} \\
  & \multicolumn{2}{c}{$(\kms)$} & 
 }

 \startdata

 2455613.668022  &   1.81 & 0.64 & HET  \\
 2455615.649694  &   2.41 & 0.64 & HET  \\
 2455616.640274  &  -2.36 & 0.64 & HET  \\
 2455623.622875  &   0.55 & 0.64 & HET  \\ 
 2455663.744361  &  -1.53 & 0.91 & Keck \\
 2455670.747785  &  -0.01 & 0.96 & Keck \\
 2455671.755651  &  -1.38 & 1.03 & Keck \\ 
 2455672.757479  &   0.10 & 1.02 & Keck \\
 2455673.770857  &   2.83 & 1.33 & Keck

\enddata
\tablecomments{RV uncertainties listed are formal
  uncertainties, and do not account for systematic effects. The offset
between the two data sets is arbitrary, and set by shifting each set so that its mean
is zero, excluding one outlier data point -- see Sections \ref{sec:HETkeck}
and \ref{sec:RVs}.}
\end{deluxetable}

%% file: discussion.tex
\section{Discussion}\label{sec:results}
\subsection{Photometry}\label{sec:photometry}

\subsubsection{Light Curves and Periodicities}\label{sec:LCandPeriods}

In order to model the transit events for a derivation of the transit
and planetary candidate properties, the effects of the
stellar variability in the light curves need to minimized -- i.e., the
light curves outside of transit needed to be whitened. After removing
data points which are flagged or have large measurement errors, we fit
a smooth cubic spline to all the PTF data which fall outside the
transit windows (using the IDL imsl\_cssmooth function\footnote{From
  the IDL Advanced Math \& Stats module; see
  \url{http://www.ittvis.com/idl}}), interpolating across the
windows themselves. The fit is then subtracted (in magnitude space) from the
entire light curve. Since most of the stellar variability occurs on
longer timescales than the transits (with the exception of the
occasional flares), this process yields the ``whitened'' light curve
that has the majority of the stellar variability removed, leaving only
the transits. Using a standalone version of the NASA Exoplanet Science
Institute (NExScI) periodogram tool,\footnote{See
  \url{http://exoplanetarchive.ipac.caltech.edu}} we calculate the
Plavchan periodogram \citep{PlavchanPeriodogram} of the combined
whitened data sets, to provide a more accurate formal transit period
measurement. We find a clear peak at
$0.448413\pm0.000040\days$.\footnote{We make the assumption that there
  is no phase shift in the transit timing between the first and second
  years, such that the datasets can be combined to give a year-long
  time baseline, and thus a very precise period estimate. If we
  disregard this assumption, the first year's data alone gives the
  best period estimate, $P=0.4486\pm 0.0010\days$.}

\begin{figure*}[tbp]
%\centering
\epsscale{1.0}
\plottwo{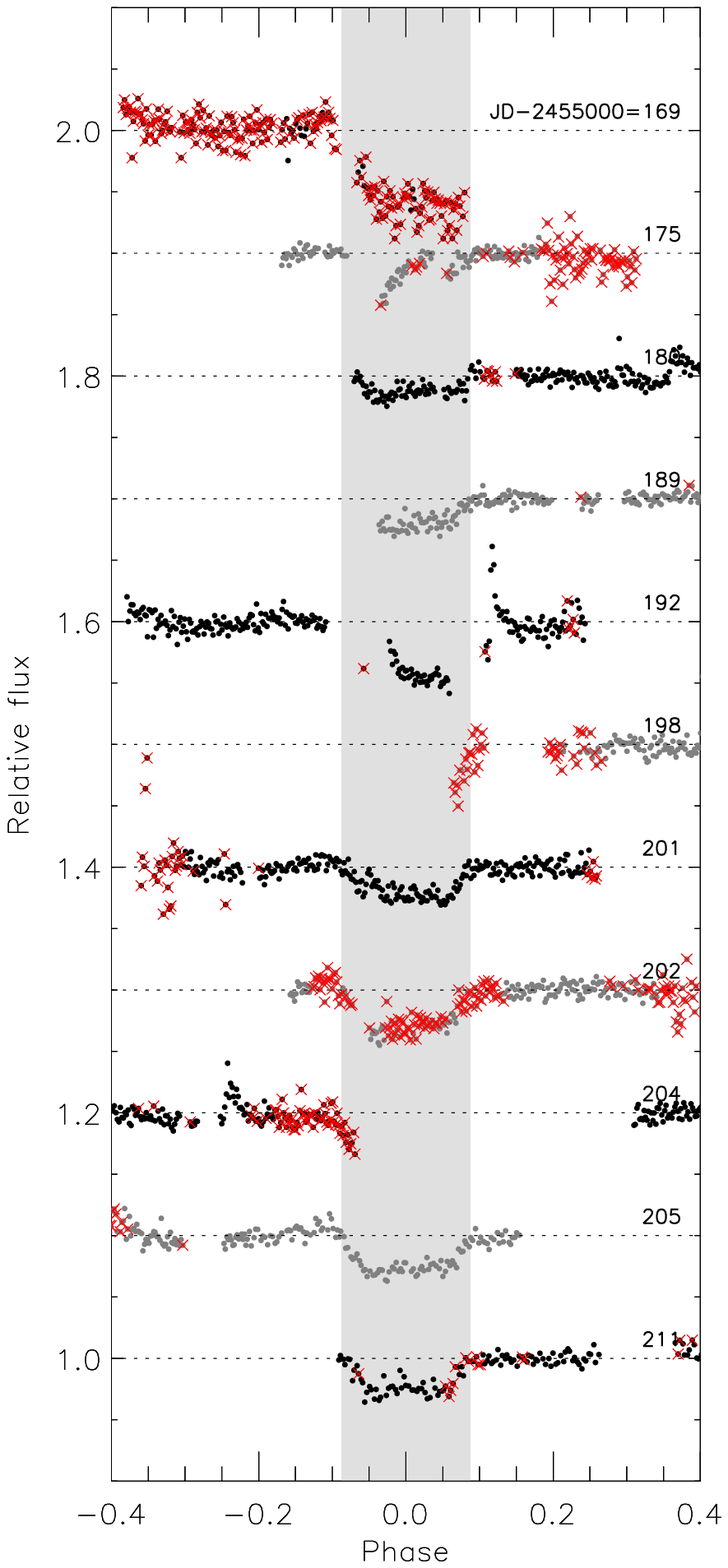}{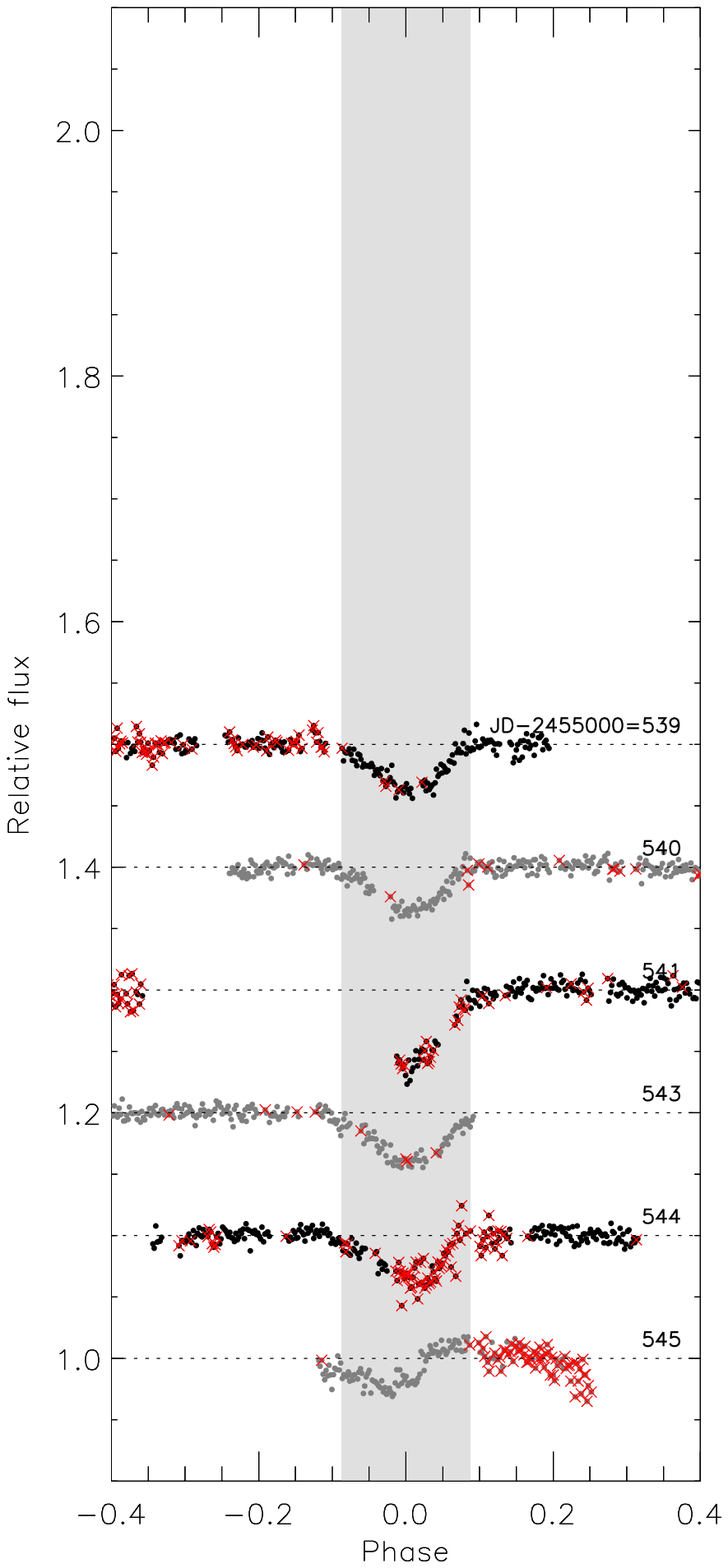}
\caption{\label{fig:foldedseparate}Whitened light curves for nights
  where in-transit data were obtained, folded on the transit period
  (Section \ref{sec:LCandPeriods}; left and right panels show first
  and second year's data respectively). Flux values are normalized to
  unity outside of eclipse, and nights are offset vertically in
  increments of 0.1 for clarity. Transits are further distinguished in
  alternating black and gray, and crosses (red in the online color
  version) indicate data
  flagged as potentially compromised. The Julian day on which each
  transit occurred is indicated for comparison with Figures
  \ref{fig:LC1} and \ref{fig:LC2}. The light gray regions indicate the
  transit windows. Note that partially covered transits with little
  or no data on one or other side of the transit window are likely to
  suffer from poor stellar-variability correction and show significant
  systematic error.}
\end{figure*}

Figure \ref{fig:foldedseparate} shows the whitened data for the nights
where any in-transit data were obtained, folded on the measured
period. Short-term stellar variability is not corrected, and is
probably the most likely explanation for those transits (particularly
those from JDs 2455175, 2455192, and 2455545) which deviate
significantly in shape from the general form of the others. This may
be caused by a combination of low-level flaring, residual disk
occultation, and the companion transiting small-scale stellar surface
features.  JD 2455192 appears to show a flare immediately after the
transit, and also likely the tail of a second flare mid-transit,
though the gaps in the data prevent a clear interpretation. JD 2455545
appears to show an early transit egress; however, the mid-transit
variation and the disparity in comparison to other transits in the
same year are suggestive of the above mentioned variability effects,
which may confuse and mask the true egress time.

\begin{figure*}[tbp]
%\centering
%\epsscale{1.15}
\plotone{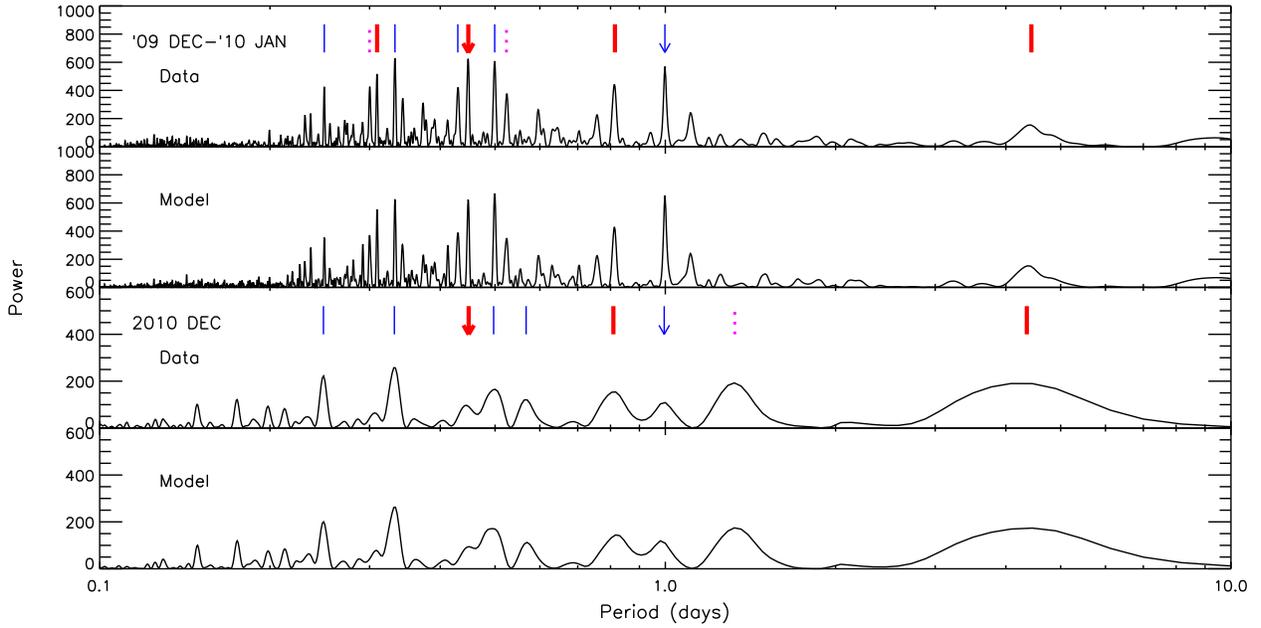}
\caption{\label{fig:LSperiodograms}Lomb-Scargle periodogram of the
  non-whitened light curves from the two years of PTF Orion
  observations, compared to simple synthetic models (Section
  \ref{sec:LCandPeriods}). The presumed stellar rotation period is
  marked with a thick arrow (red in the online color version); a
  second distinct one-day period, which we attribute to the observing
  cadence, is marked with a thin (blue) arrow. The more prominent
  aliases of these periods are indicated with vertical lines: thick
  (red) marks indicate peaks which are predominantly produced by
  aliases of the presumed stellar rotation period; thin (blue) marks
  indicate peaks predominantly produced by aliases of the one-day
  period. Dotted (magenta) lines indicate peaks produced by a
  combination of the two.}
\end{figure*}

The original ``un-whitened'' light curve is dominated by stellar
variability (see Figures \ref{fig:LC1} and \ref{fig:LC2}). If we
assume that the large-scale variability is caused primarily by spot
modulation, and that the relative effect of the transit events is
negligible, we can use the un-whitened data to investigate the stellar
rotation. Lomb-Scargle periodograms \citep{Lomb1976, Scargle1982} of
the two years' data are shown in Figure \ref{fig:LSperiodograms}. Where
the Plavchan periodogram is designed for finding regular periodic
features of arbitrary but constant shape (such as a transit), a
Lomb-Scargle periodogram is better suited for finding periodic modes
in a more complex signal such as the quasi-periodic stellar variation
in our data: here, the signal appears not to repeat in
an exact fashion and therefore does not fold well at any period
(likely because of phase shifts due to changing spot features).

A strong peak is found at $0.4481 \pm 0.0022\days$, in fact matching
well with the transit period. Another peak is seen at $0.9985 \pm
0.0061\days$, corresponding closely to the sidereal or solar Earth
day, and therefore most likely an artifact resulting from the
observing cadence. The other peaks all appear to be aliases of these
two periods. To confirm this, we modeled the data by creating an
artificial light curve from two summed sinusoids with the same two
periods, providing power at the frequencies of these two peaks. We
superimposed artificial transits modeled as a simple inverted top-hat
function, with the same depth, width, and ephemeris as the real
transits. Allowing the phases and relative amplitude of the
$0.4481\days$ and $0.9985\days$ signals to vary as free parameters,
and requiring that the total amplitude of the light curve remain
approximately the same as that of the actual data, we performed a
least-squares fit to the real periodogram to assess how well its
structure could be reproduced. Figure \ref{fig:LSperiodograms} shows
the results, with a
good match between model and data. Repeating the test with the
$0.4481\days$ sinusoid omitted (i.e., summing only the one-day signal
and the transits) gave a poor fit, implying that the form of the
periodogram cannot be explained by the transits alone. 
Since there is clearly strong correlated out-of-transit variability
associated with the star, and there are no other
fundamental periods evident in the periodogram, the $0.4481\days$ signal
appears to be the only likely period for the star.

There are three other notable aliases of the presumed stellar rotation
period, $P_*$, at $0.3092 \pm 0.0010\days$, $0.8126 \pm 0.0070 \days$, and
$4.43 \pm 0.31\days$ (identified in the figure). The larger of these
is substantially above the upper period limit implied by the measured
$v_\mathrm{r}\sin i_*$ of the star (Section \ref{sec:vsini}), $P_* <
2\pi R_* /(v_\mathrm{r}\sin i_*) = 0.671 \pm 0.092\days$. Though it is
unlikely that the star would be coincidentally rotating at an alias of
the transit period with the Earth's rotation, we repeated the modeling
experiment with the model stellar rotation modified to match the
remaining two aliases to ensure that none could in fact be the true
fundamental stellar period \citep[see][]{Dawson2010}. Similar
periodograms were obtained, with peaks at similar locations but
with differing strength ratios; none were as good a fit to the periodograms
based on the real data. We therefore conclude that the star is
co-rotating or near co-rotating with the companion orbit.

We note that it is difficult to model the observed transit events with
star spots alone, particularly in the 2009 December/2010 January data:
the short transit duration relative to the period, the flat bottom of
the transits, and the sharp ingress and egress, are all more
characteristic of a transiting object, whereas spots tend to cause
smoother, more sinusoidal features as their projected area changes as
they rotate across the stellar disk. It is more likely that we are
seeing in the total lightcurve the effects of both star spots and a
transiting object.

\subsubsection{Transit Fitting}\label{sec:transitfitting}
\begin{figure}[tbp]
%\centering
\epsscale{1.0}
\plotone{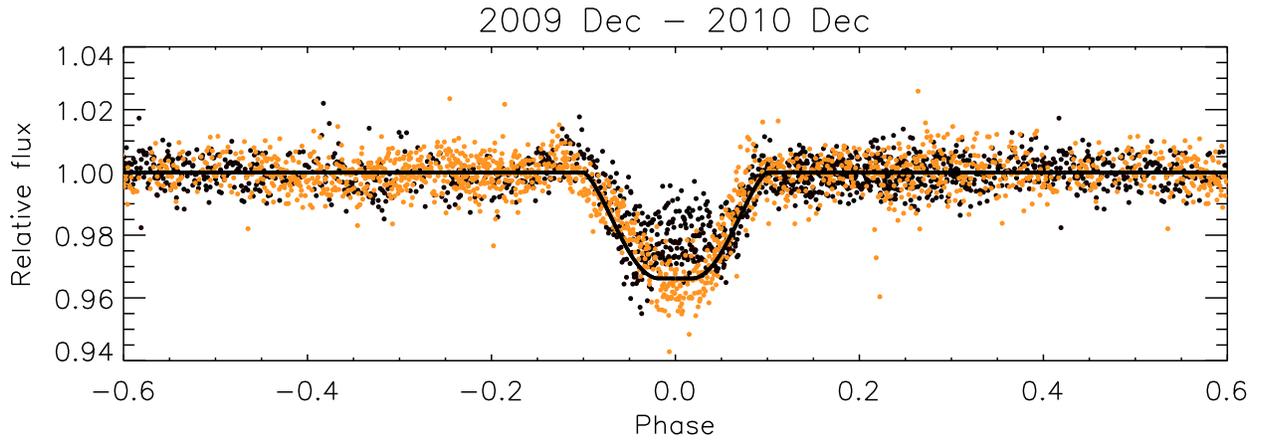}
\caption{\label{fig:transitfit} Folded light curve for the two years
  of observations combined, after removing stellar variability
  (Section \ref{sec:LCandPeriods}). The first year's data (2009
  December -- 2010 January) are plotted in black; the second year's
  (2010 December) are plotted in orange. A change in transit shape
  between the two years is evident. The best transit fit to both years
  combined is over-plotted on the assumption that $R_p < R_*$ (Section
  \ref{sec:transitfitting}). Limb darkening is neglected, and error
  bars are omitted for clarity; the median photometric error is
  0.0046, with an out-of-eclipse standard deviation of 0.0056.}
\end{figure}

We fit transit models to the whitened light curves using the IDL-based
Transit Analysis Package \citep[TAP;][]{Gazak2012}. TAP employs Bayesian Markov Chain
Monte Carlo (MCMC) techniques to explore the fitting parameter space
based on the analytic transit light curve models of
\citet{Mandel2002}. The package incorporates white- and red-noise
parameterization \citep{Carter2009}, allowing for robust estimates of
parameter uncertainty distributions. We fit only the transits from
nights where a complete transit was observed with adequate coverage
both before and after the transit window, since on these nights the
spline fit to remove the stellar variability is reasonably constrained
on both sides of the transit; on nights where there is partial transit
coverage, the spline fit tends to diverge during transit time. We also
reject the transits mentioned in section \ref{sec:photometry} where
stellar variability appears to strongly affect the shape of the
transit, and in addition, the transit from JD 2455544, where the
spline subtraction was particularly uncertain and rather sensitive to
the tightness of the spline fit. Thus we fit four of the transits
from the first year's data (JDs 2455201, 2455202, 2455205, and
2455211), and three from the second year (JDs 2455539, 2455540, and
2455543). The folded data are shown in Figure
\ref{fig:transitfit}.

For the transit fitting, we hold the orbital period, $P$, fixed
to the transit period determined above. The eccentricity,
$e$, is fixed at zero, since it is difficult to constrain in the
absence of a secondary eclipse. White and red noise levels are allowed
to float separately for each individual transit, as are linear airmass
trends. The remaining parameters are allowed to float, but are tied
across all transits. They include: the epoch of transit center, $T_0$
(we assume there are no transit timing drifts between the two years);
the orbital inclination, $i_\mathrm{orb}$; $a/R_*$, where $a$ is the
orbital semi-major axis, and $R_*$ is the radius of the primary;
$R_\mathrm{p}/R_*$, where $R_\mathrm{p}$ is the companion radius; and
the linear and quadratic limb-darkening coefficients for the primary
star. In addition, we constrain the fitting such that
$R_\mathrm{p}/R_* < 1$, and white and red noise components are less
than 4\% (the depth of the transit). Parameters are calculated by
fitting Gaussians to the dominant peak in the probability density
distributions for each parameter resulting from the MCMC analysis. The
center location of the Gaussian is taken as the parameter value, and
measurement errors are estimated as the dispersion of the fit. We find
that the limb-darkening coefficients are largely unconstrained by the
data. The fit is shown overlaid on the folded data in Figure
\ref{fig:transitfit}; the corresponding parameters are listed in Table
\ref{tbl:transitparams}.

\input{table-transitparams}

There is significant variation in the light curve from transit to
transit, as can be seen in Figure \ref{fig:foldedseparate}. This can
reasonably be attributed to the companion passing across varying
surface features -- cold or hot spots, or perhaps flares -- on the
stellar photosphere. Such variations are a result of the planet
tracing the stellar surface brightness profile as it traverses the
stellar disk, and cannot be removed by the whitening process, which is
only sensitive to the integrated brightness of the disk. Since PTFO
8-8695 is expected to be active and spotted, such variation in the
transits is further suggestive of a genuine transit, rather than a
background blend.

It can also clearly be seen from Figures \ref{fig:foldedseparate} and
\ref{fig:transitfit}, however, that there is an overall change in the
transit shape between the two years' data sets. Explanations for this
remain speculative. Re-running the fitting process and allowing a
change in both $R_*$ and $R_\mathrm{p}$ between the two years yields a
decrease in stellar radius of $\approx10$\% $(2.1\sigma)$ from one
year to the next (with no measurable change in companion radius). 
Such a large change in the stellar radius in such a short time period is unlikely, and would likely manifest itself in observable rotation rate changes not seen in the data. A change in transit shape
could also arise in principle from a change in orbital geometry. For
example, the host star is rotating quickly enough to exhibit
significant oblateness, which may induce precession of the orbital
plane and, therefore, changes in $i_\mathrm{orb}$ if the orbital and
stellar rotational axes are misaligned.\footnote{Given the small
  orbital period, this may occur on short timescales. Following
  \citet{MiraldaEscude2002}, we estimate the order of magnitude of the
  oblateness-induced gravitational quadrupole moment of the host star,
  $J_2$, by scaling to the pre-main--sequence from a solar value of
  $\sim 10^{-7}$--$10^{-6}$ according to $J_2 \propto R_*^3/(M_*
  P_*^2)$. Assuming that the stellar obliquity is small, and noting that
  the orbital angular momentum and stellar rotational angular momentum
  are similar for $M_\mathrm{p}\sim 5\Mjup$ (so that the inclinations
  of the stellar rotation and orbital planes with respect to
  the mean plane are similar), this leads to a
  precession period of the orbital nodes on the order of tens to
  hundreds of days.}
An additional planet in a different orbit could also produce a similar
effect \citep{MiraldaEscude2002}. It is difficult, however, to explain
how an inclination change can yield a transit that becomes both more grazing (longer ingress/egress) and
deeper at the same time. Another explanation may be
variation in star spot coverage (or limb-darkening, though with heavy
spot coverage, the two effects become somewhat confused). In addition
to short-term spot variations, there may be surface features which
survive for much longer periods \citep[see, e.g.,][]{Mahmud2011}. Given the
activity of the star and the extreme proximity of the companion, such
features may be compounded by magnetic or tidal star/planet
interactions that could give rise to varying hot or cold spots near
the sub-planetary point on the stellar photosphere. This could affect
the apparent shape of the transit in a systematic way. The planet's
apparent proximity to the tidal disruption limit (see section
\ref{sec:limits}) may also be a factor: a tidally distorted shape, or
transient rings or a tidal tail of evaporating material could all
yield unexpected and possibly varying transit shape (and also cause
the slight transit asymmetry seen in the second year's data).  For the
sake of simplicity, and given the lack of data to disentangle all of
these possibilities, we here adopt the single combined fit to both
years' data sets, although with the caveat that the variability may
cause some systematic error in the measurements.

We note that our fit yields a somewhat smaller stellar radius,
$R_*\approx1.07\Rsun$, than that previously reported by
\citet{Briceno2005} ($1.39\Rsun$). Assuming their estimate of
$T_\mathrm{eff}=3470\K$ is correct, interpolating Siess models
\citep{SiessModels} with this updated radius gives a slightly older
age estimate for the star, $\approx 3.7\Myr$ vs. $2.7\Myr$. Given the
distance uncertainties in the \citet{Briceno2005} results, however,
and the possibility of uncertainty in $T_\mathrm{eff}$ arising from
heavy spotting in the stellar photosphere, the two radius estimates
are probably not inconsistent.\footnote{Our radius estimate also
  depends on our estimate of $a$, which depends in turn on the
  assumption that the mass estimate of \citet{Briceno2005} is correct,
  so the argument is somewhat circular. $a$, however, is relatively
  insensitive to stellar mass ($a\propto M^{1/3}$), and the Siess
  models predict a negligible difference in mass at our older age. (In
  fact, at the estimated $T_\mathrm{eff}$, they predict a mass range
  smaller than the errors over an age range of 1--$10\Myr$.) }

\subsection{Spectroscopy}

\subsubsection{Radial Velocities}\label{sec:RVs}
\begin{figure}[tbp]
%\epsscale{1.15}
\plotone{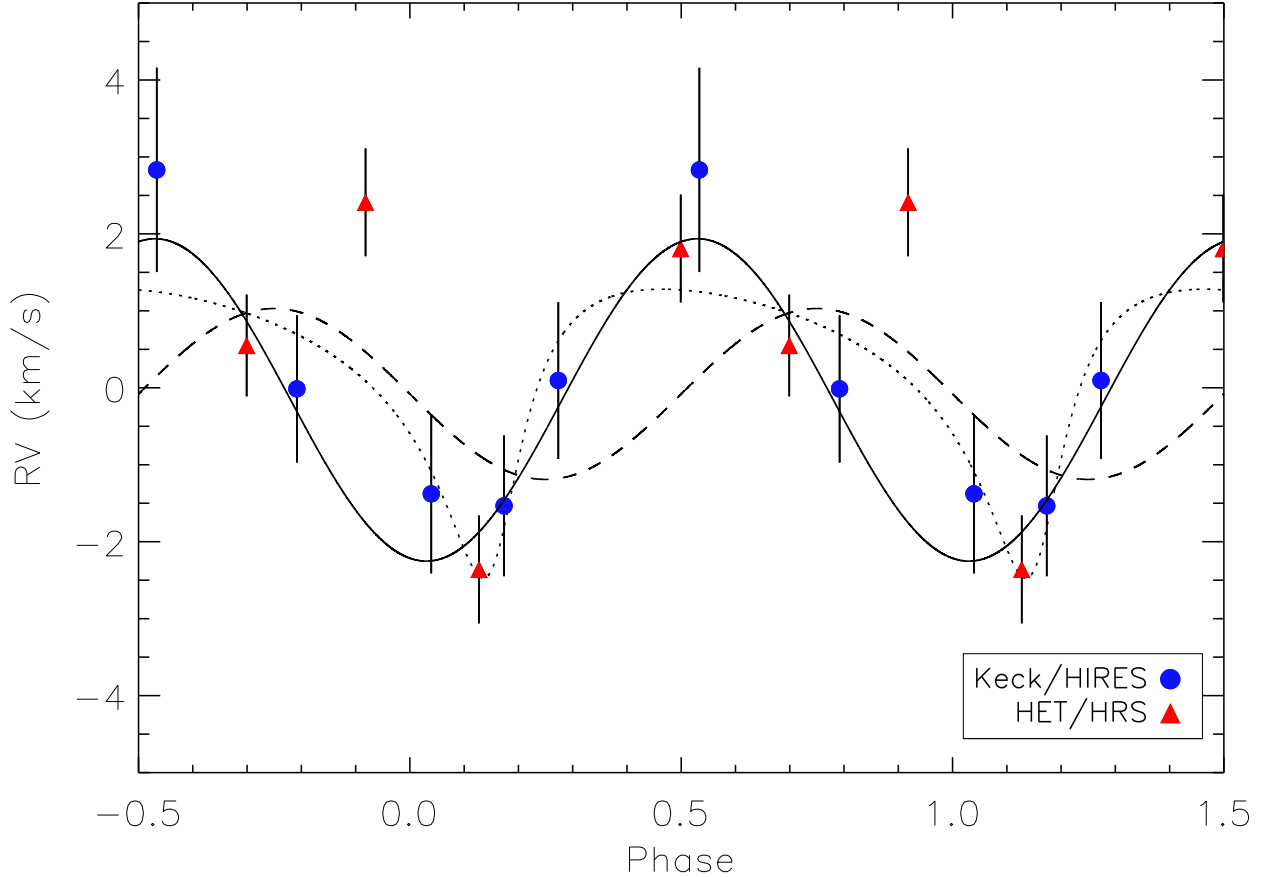}
\caption{\label{fig:RVs} Differential radial velocity measurements
  obtained with Keck/HIRES and HET/HRS (Section \ref{sec:RVs}). Zero
  phase is chosen to correspond to the photometric center-of-transit
  time. The offset between the two datasets is chosen to match the
  mean RV of each. The lines indicate best Keplerian fits (excluding the
  outlier): dashed line -- circular orbit, transit-center time, $T_0$,
  fixed to photometry ($\rchisq=4.0$); dotted line -- eccentric orbit,
  transit-center time fixed to photometry ($\rchisq=1.1$); solid line
  -- a sinusoidal fit (equivalent to a circular orbit) at the same
  period, with phase free to float ($\rchisq=0.42$). The eccentric fit
  is better than the fixed-$T_0$ fit, but brings the companion to the
  surface of the star at periastron. The floating-phase sinusoidal fit
  gives the best result, suggesting that star spots either modify or
  dominate the Doppler RV signal.  The outlier point may represent
  Rossiter McLaughlin effect due to the transiting companion.}
\end{figure}

Since the RV analysis is differential in nature, the RV offset between
the HET and Keck data sets is arbitrary. We place them on the same
approximate scale by shifting the RV zero points to the mean of the
data for each data set. There are too few data points to create a
periodogram to measure any periodicities in the data, but we can look
for consistency with the previously determined transit period by
folding the data on that period and looking for a coherent alignment
of the data points. The result is shown in Figure \ref{fig:RVs}:
indeed, the data appear to phase well, showing a smooth and apparently
sinusoidal variation, with the exception of one outlying data point at
orbital phase $\phi=-0.082$ (Heliocentric Julian date (HJD)
2455615.64969), which is discussed further below. This outlier is
excluded in calculating the offset between the data sets.

To constrain the mass of the companion, we fit a Keplerian orbit
model to the data using the RVLIN package by \citet{RVLIN}. The model
includes six parameters: the period, $P$; the RV semi-amplitude, $K$;
the eccentricity, $e$; the argument of periastron, $\omega$; the time
of periastron passage, $T_\mathrm{p}$; and the systemic velocity, $\gamma$.
We fixed $P$ and $T_\mathrm{p}$ to the values measured from the transit
photometry.  Since the $\chi^2$ surface has multiple minima, we
explore the fitting parameter space by running 10000 trial Keplerian
fits (neglecting the apparent outlier point), with initial parameter
estimates drawn at random from reasonable starting distributions, and
selecting the fit with the lowest reduced $\chi^2$ ($\rchisq$).

We find that the constraints from the transit ephemeris make it
difficult to obtain a reasonable Keplerian fit. A good circular-orbit
fit, with the eccentricity, $e$, fixed to zero, is prohibited by the
constraint on $T_0$ provided by the photometry. Such a fit
must cross its central velocity at phases 0 and 0.5, with minima and
maxima at 0.25 and 0.75 (the quadrature points), and as a result
appears significantly out of phase with the data ($\rchisq=4.0$;
dashed line, Figure \ref{fig:RVs}).

An eccentric fit goes some way to resolving the mismatch
($\rchisq=1.1$), but yields an eccentricity of $e\approx 0.492$ (dotted line,
Figure \ref{fig:RVs}). This places the fractional periastron distance
at $r_\mathrm{peri}/a \equiv 1-e \approx 0.51$, bringing the companion
to the point of contact with the surface of the star as determined
from the photometric transit fits ($R_*/a \approx 0.51$). Such a
solution may be improbable, given that in section \ref{sec:limits} we note that
the orbital semi-major axis appears to be close to the Roche
limit.

A better fit is in fact obtained by fitting a circular orbit model and
removing the constraint on $T_0$, allowing the phase to float.
This fit is over-plotted in Figure \ref{fig:RVs}, showing the fit is
good ($\rchisq=0.42$), but offset in phase from the transit ephemeris
by $-0.22\pm 0.04$ periods. 

We favor this better fit to the RV signal, which most likely arises
because of spot effects modulated by the stellar rotation, where the
amplitude of the spot effect is at least comparable to -- if not much
greater than -- any true reflex Doppler signal from the
companion. Similar spot-induced RV amplitudes have already been
observed for other T-Tauri stars in the optical
\citep[e.g.,][]{Mahmud2011,Huerta2008,Prato2008}. Since the star
appears to be co-rotating with the planet orbit, the period of such a
spot-dominated RV signal would match the orbital period, but the phase
would be arbitrary, depending on the longitudinal spot
placement. Alternatively, spot and companion RV signals may both be
significant and the two effects may compete (see Section \ref{sec:limits}).

Regardless of the astrophysical cause behind the measured RV signal, we
can place an upper limit on the companion mass, and because of the good orbital phase
coverage we can be reasonably assured that possible aliasing effects
are not a concern. We assume that any companion-induced
Doppler motion must be smaller than the amplitude of the measured
variations, that $M_p \ll M_*$, and that $M_*=0.39 \pm 0.10
\Msun$ \citep{Briceno2005}, and hence estimate $M_\mathrm{p} \sin
i_\mathrm{orb} \le 4.8\pm 1.2\Mjup$ \citep[see, e.g., formula
3,][]{Gaudi2007}. Taking our derived value of $i_\mathrm{orb}$ we can
directly estimate $M_\mathrm{p} \le 5.5 \pm 1.4\Mjup$, comfortably
within the planetary mass regime. The semi-amplitude (half
peak-to-peak) of the combined measured RV variations is $\approx
2.4\kms$. This is separately confirmed by the two individual
datasets, which both have good phase coverage and show similar
amplitudes independent of the RV offset (as can be seen in Figure
\ref{fig:RVs}). By comparison, since Doppler-induced RV semi-amplitude
scales directly with $M_\mathrm{p}$, a $25\Mjup$ object on the
planet/brown-dwarf boundary suggested by \citet{Schneider2011} would
induce an $11\kms$ signal; an object on the deuterium-burning limit
($\approx 13\Mjup$) would induce a $5.7\kms$ signal. It is unlikely,
therefore, that the object is of more than planetary mass, and still
less in the stellar mass range.\footnote{We do note, however, that
  there is a possibility that if the reflex Doppler signature and the
  spot signature are of comparable amplitudes, matching periods, and
  opposed in phase, they may destructively interfere, giving an
  artificially low RV amplitude.} Since the spot distribution is
likely to change with time, further observations to look for changes
in the phase of the RV signal with respect to the transit ephemeris
may provide more insight into the exact nature of the RV signal. RV
observations in the infra red are also likely to suffer less from
spot-induced noise, and so could also provide valuable further
information.

We were unable to find any abnormalities in the observations regarding
the outlying RV data point, and, thus, have no reason to reject it as
a bad measurement. However, falling at an orbital phase $\phi=-0.082$,
it lies within the transit window (c.f. Figure \ref{fig:transitfit}),
and its anomalous velocity could reasonably be explained as arising
from the Rossiter-McLaughlin (RM) effect
\citep{Rossiter1924,McLaughlin1924}. The RM effect would appear
regardless of whether the main RV signal is caused by spot modulation
or stellar reflex motion, since it is a result of an asymmetric
distortion of the stellar absorption lines as an obscuration transits
the unresolved stellar disk, blocking off regions with successively
different rotational red-shifts. Indeed, given the rapid stellar
rotation, the RM effect {\em should} be expected to appear during the
transit window. From equation 6 of \citet{Gaudi2007} we can estimate
the expected maximum amplitude of the effect, $K_R$, given
$v_\mathrm{r}\sin i_*$, $R_p$, and $R_*$.  We find $K_R\approx 3\kms$,
in good agreement with the offset of the outlier. The RM maximum also
lies typically around $\sim1/5 $--1/3 of the way between
transit-ingress and transit-center for gas-giant planets, which is
coincidentally where the outlier point is located. The sign of the
offset of the outlier data point, falling prior to the transit center
time, is in agreement with a prograde orbit, as we would expect if the
star is in synchronous (or quasi-synchronous) rotation. It should be
noted that the RV data point at $\phi=+0.039$ (HJD 2455671.755651)
also lies within the transit window but, in contrast, does not appear
to be an outlier. Lying closer to the transit center time, one would
expect the RV offset to be smaller here; however, the exact form of
the RM effect depends on the precise orbital geometry, and may be
partially masked by the choice of offset between the two data
sets. Further complication may also arise from the companion
transiting photospheric surface features. Dedicated RV measurements
during transit could provide valuable confirmation of the RM effect
hinted at by the data, and help independently confirm the validity of
the transiting planet candidate.

\subsubsection{Stellar Rotation}\label{sec:vsini}

The stellar rotation velocity, $v_\mathrm{r}$, measured from the
spectroscopy provides an independent consistency check on the stellar
rotation period. In order to estimate the projected rotation velocity,
$v_\mathrm{r}\sin i_*$, 
we degrade the Kurucz synthetic model spectrum (see Section
\ref{sec:HETkeck}) to the resolution and sampling of the HRS, and
rotationally broaden it using a non-linear limb-darkening model
\citep{Claret2000, Gray1992} over the range of $v_\mathrm{r}\sin i_* =
10$ -- $150\kms$.  We then Doppler shift the broadened models, line by
line, by cross-correlating against lines in our target spectrum.
Subtracting a Doppler shifted model from the observed spectrum gives
an RMS residual for that rotational velocity.  Minimizing this
residual gives the optimal rotational velocity measured for an
individual line.  We applied this analysis to 12 lines that were
sufficiently deep to be visible above the noise for our full range of
$v_\mathrm{r}\sin i_*$ models.  Weak lines could not be used because
at large $v_\mathrm{r}\sin i_*$ the line profiles became buried in the
noise.  The weighted average of $v_\mathrm{r}\sin i_*$ measured for
these lines, across all four HRS observations, was $80.6\pm8.1\kms$.

The assumed rotational period of the star and the radius measured from
photometric transit fitting (neglecting oblateness effects)
independently imply an expected value for the unprojected rotational
velocity of $v_\mathrm{r} = 2\pi R_*/P_* \approx 120\pm 11\kms$. Taken
together, the measured $v_\mathrm{r}\sin i_*$ and the photometrically
derived $v_\mathrm{r}$ give an estimated inclination of the stellar rotation
axis, $i_* = 42\pm 7\degr$. Comparing this with the measured orbital
inclination, $i_\mathrm{orb} = 61.8\pm 3.7\degr$, we see that there is
weak evidence for a possible misalignment between the orbital and
stellar rotation axes (consistent with the possibility of detecting
changes in the orbital inclination on relatively short timescales, as
mentioned in Section \ref{sec:transitfitting}).

\subsection{Implications}\label{sec:limits}

\begin{figure}[tbp]
%\centering
%\epsscale{1.0}
\plotone{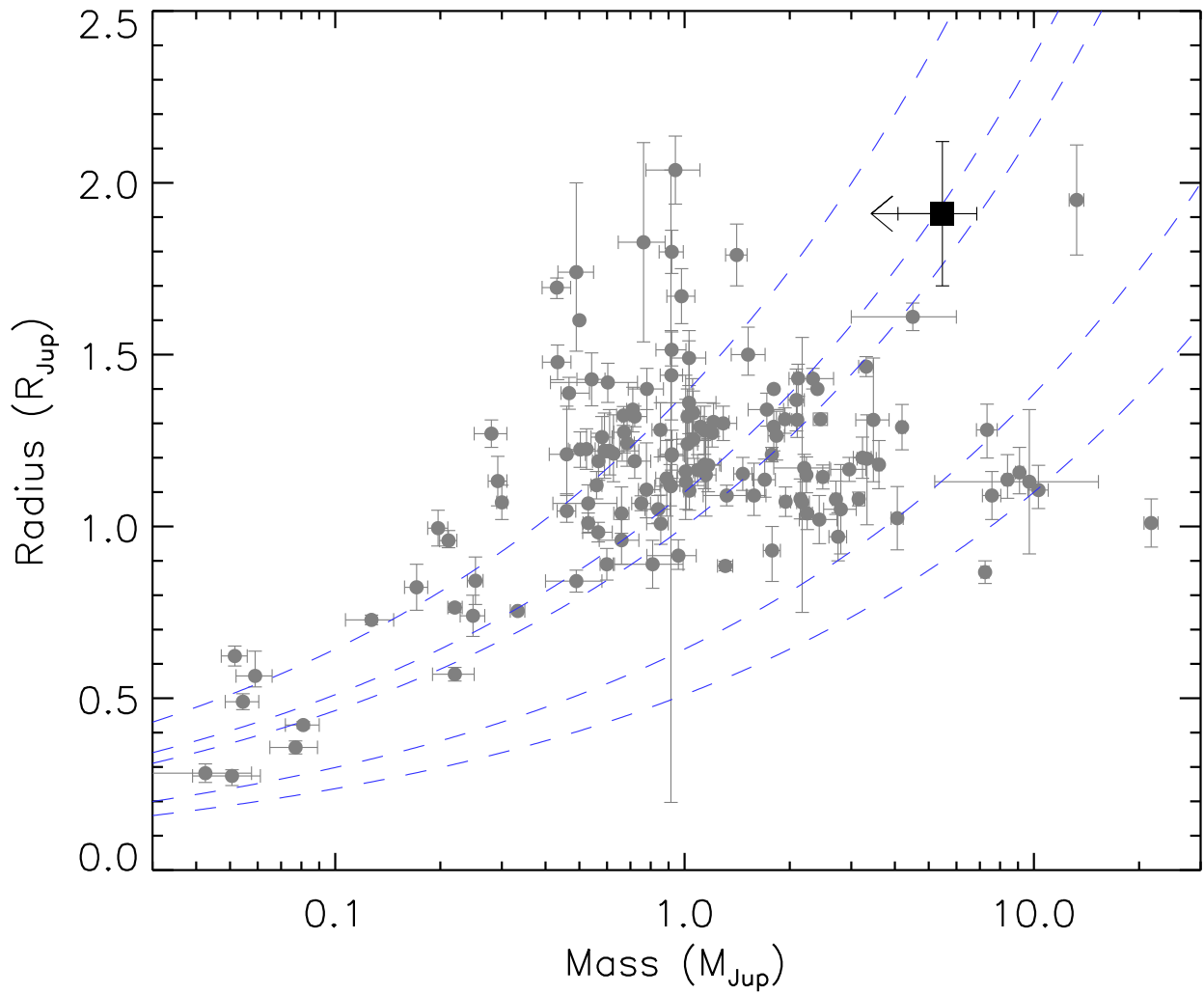}
\caption{\label{fig:density} Radius vs.\ mass for the known transiting
  planets (Section \ref{sec:limits}; data taken from the NASA
  Exoplanet Archive). PTFO 8-8695b is marked as an upper mass limit,
  highlighted by the large square. Iso-density contours are marked at
  0.5, 1, 1.33 (Jupiter density), 5, and $10\mathrm{\, g\,cm^{-3}}$.}
\end{figure}

In Figure \ref{fig:density} we show the radius and mass of the
companion in relation to the currently known transiting exoplanets
listed in the NASA Exoplanet
Archive,\footnote{\url{http://exoplanetarchive.ipac.caltech.edu}}
where we have indicated the mass of the candidate as an upper limit
only. It clearly lies at the upper end of the gas-giant radius
distribution, although not unprecedentedly so. A significantly inflated
atmosphere is to be expected owing both to the system's very young
age, and to stellar irradiation at the companion's exceptionally small
orbital radius. Planetary atmosphere models are currently not well
constrained in this regime due to the lack of known exoplanets that
are both young {\em and} close-in. \citet{Marley2007} and
\citet{Fortney2007} caution that evolutionary models should be treated
with care at ages up to $10\Myr$ or more, being highly dependent on
initial conditions and the formation mechanism assumed. Indeed,
\citet{Spiegel2012} suggest that comparison of atmospheric models with
observations of exoplanets at such young ages may be a good way of
distinguishing between different formation scenarios.
Both \citet{Spiegel2012} and \citet{Marley2007} predict radii of
around $1.6\Rjup$ for a $5\Mjup$ planet at $3\Myr$, for non-irradiated
`hot-start' (gravitational collapse) model atmospheres; our error bars
place our measured radius ($1.9\pm0.2\Rsun$) just above this
value. Our radius lies substantially further above the post-formation
cold-start (core collapse) isochrones, though comparison with the
latter models is confused by uncertainty in the formation timescale,
which is likely comparable to the age of the star. Accounting for
stellar irradiation could increase the theoretical radius
further. Increases of $\sim 10\%$ or more are typical at $\sim
0.02\AU$ from a solar like star \citep{Baraffe2010,Chabrier2004},
where the incident flux is similar to that on our planet candidate
(which is closer in, but orbiting a lower-mass star). The companion's
proximity to the Roche limit may also weaken the gravitational binding
of the object, further increasing its expected radius. Various other
explanations are also proposed to explain the excess ``radius
anomaly'' that is increasingly found in other gas-giant exoplanets
\citep[e.g.,][]{Baraffe2010,Chabrier2011}. Our estimated radius is
therefore not unreasonable, but due to both the model and observation
uncertainties, a robust comparison with theory is difficult.

The apparent companion to PTFO 8-8695 also orbits close enough to its parent star that
the consequent small size of its Roche lobe may be relevant. Below a certain
mass threshold, it may not have sufficient self-gravity to hold
itself together, and thus may begin to lose mass. Following \citet{Faber2005} and
\citet{Ford2006}, the Roche radius, $R_\mathrm{Roche}$,  according to
\citet{Paczynski1971}, is given by 
\begin{equation}
R_\mathrm{Roche} \; = \;
0.462(M_\mathrm{p}/M_*)^{1/3}a \; 
\le \; 0.462(M_\mathrm{p,max}/M_*)^{1/3}a \;
 \approx \; 1.92 \pm 0.16,
\end{equation}
where $M_\mathrm{p,max}$ is our previously estimated maximum mass for
the companion. Comparing with the measured companion radius gives
$R_\mathrm{p}/R_\mathrm{Roche} \gtrsim 0.994\pm 0.094$.  Framing the
argument another way, we can rewrite the Roche formula to estimate the
Roche limiting orbital radius, $R_\mathrm{Roche}$, in terms of the
measured companion radius to find $a/a_{\mathrm{Roche}} \lesssim
1.008\pm0.095$ at the most -- consistent with being at or within the
Roche limit, within the errors. The planet may be sufficiently
inflated that it fills its Roche lobe, and consequently may have lost mass
in the past, or be in the process of losing mass (thus maintaining
itself at the Roche limit).

%% file: table-transitparams.tex
\begin{deluxetable}{rl}
%\tabletypesize{\scriptsize}
\tablecaption{System Parameters\label{tbl:transitparams}}
%\tablewidth{0pt}
\tablehead{\colhead{Parameter} & \colhead{Value}}
\startdata
\sidehead{\bf{Measured}}
$P$                    & $0.448413 \pm 0.000040\days$ \\
$i_\mathrm{orb}$                    & $61.8 \pm 3.7\degr$\\
$a/R_*$                & $1.685 \pm 0.064 $ \\
$R_\mathrm{p}/R_*$      & $0.1838 \pm 0.0097 $ \\
$T_0$ (HJD)            & $2455543.9402 \pm 0.0008 $ \\
$v_\mathrm{r}\sin i_*$              & $80.6 \pm 8.1 \kms $ \\
\sidehead{\bf{Derived}}
$a$                    & $0.00838 \pm 0.00072\AU $\\
                       & $ = 1.80 \pm 0.15\Rsun $\tablenotemark{a}\\
$R_*$                  & $ 1.07 \pm 0.10 \Rsun $\\
$R_\mathrm{p}$          & $ 1.91 \pm 0.21\Rjup$\\
$M_\mathrm{p} \sin i_\mathrm{orb}$   & $\le 4.8 \pm 1.2\Mjup$\tablenotemark{b}\\
$M_\mathrm{p}$ & $\le 5.5 \pm 1.4\Mjup$\\
\enddata
%Min $M_\mathrm{p}$ & $5.34 \pm 1.72\Mjup$\tablenotemark{c}\\
\tablecomments{Summary of parameters determined in this paper for the
  PTFO 8-8695 system. Quantities are $P$ -- orbital period;
  $i_\mathrm{orb}$ -- orbital inclination; $i_*$ -- inclination of stellar
  rotation axis; $a$ -- orbital semi-major axis; $R_*$ -- stellar
  radius; $R_p$ -- planet radius; $T_0$ -- epoch of transit center;
  $v_\mathrm{r}$ -- stellar equatorial rotational velocity;
  $M_\mathrm{p}$ -- planet mass.}
\tablenotetext{a}{From Kepler's third law, assuming stellar mass $M_*=0.39\pm 0.10\Msun$
  \citep{Briceno2005}, and $M_\mathrm{p}\ll M_*$.}
\tablenotetext{b}{Upper limit derived from measured RV
  semi-amplitude.}
  %Roche limit considerations suggest a lower limit
  %close to this value, $M_\mathrm{p} \gtrsim 5.3\pm1.7\Mjup$ (see
  %section \ref{sec:limits}).}
\end{deluxetable}

%% file: conclusions.tex
\section{Conclusions}\label{sec:conclusions}

We have detected transits from a candidate young planet orbiting a
previously-identified co-rotating or near-co-rotating $\approx
2.7\Myr$-old M3 weak-line T-Tauri star. Although we cannot completely
rule out a false positive due to source blends, we are able to rule
out confusion at the level of $\Delta H \approx 4.3\vmag$ beyond a
separation of 0.25\arcsec, and argue qualitatively that a false
detection due to a blended eclipsing binary is unlikely. The companion
is in an exceptionally rapid $0.448413\days$ orbit, placing it among
the shortest of the currently known exoplanet periods \citep[cf.][]{Demory2011,
Charpinet2011, Muirhead2012, Rappaport2012}.
With an orbital radius only around twice the stellar radius, it
appears to be at or within its Roche limiting orbit, with
$a/a_\mathrm{Roche}\lesssim 1.01 \pm 0.10$, raising the possibility of
past or ongoing evaporation and mass loss. Perhaps the companion has
been migrating, and losing any mass beyond its Roche lobe as it
does so; or perhaps it is continually being inflated to fill its Roche
lobe, with any material which overflows being stripped away.

Although the transit photometry and the RV data both phase-fold on the
same periods, there is an apparent offset in phase between them. The
most likely explanation is that the RV signal is shifted or dominated
by the effect of star spots; we therefore suggest an upper limit on
the (inclination independent) companion mass of $\approx 5\Mjup$ based
on the amplitude of the RV modulation. If it can be assumed that the
object has had time to reach a stable (or quasi-stable) state
 -- i.e., that mass loss rates are not too rapid, and that there have been
no recent dramatic changes in the orbital geometry -- then Roche
limit considerations would imply a lower limit of a similar
order, since anything much less would be unable to gravitationally bind
the material within the measured companion radius.

The data are complex enough that we cannot yet be certain of a planet
detection. In favor of the planetary interpretation, however, we note
that: (1) the photometric transit shape appears to be flat bottomed
with sharp ingress and egress slopes, and is difficult to explain with
a star-spot model alone; (2) the transits appear highly consistent and
periodic over a period of $\gtrsim 1\,\mathrm{yr}$, which would be
unusual for spots; (3) the RV signal places an upper mass limit well
within the planetary regime; (4) a false positive due to a faint
blended eclipsing binary is argued against by two observations which
are suggestive of a transiting object associated with the primary
T-Tauri star observed: the stellar rotation rate appears very close to
the transit period, suggesting co-rotation; and the photometric
variation from transit to transit is consistent with the notion of a
planet transiting a heavily spotted T-Tauri star.

Further spectroscopic and photometric observations are needed to
provide a clearer picture. Since the star is such a fast rotator, the
expected RM effect (hinted at in the data) may provide a valuable opportunity for
confirmation of the candidate. Most RV noise sources will be constant
in the timescale of one transit, so further RV observations during the
transit window could potentially provide full sampling of the
effect. This would help confirm the planetary interpretation of the
data, and could provide further useful information on the system
geometry. Revisiting the target to make further measurements of the RV
phase with respect to the transit ephemeris would help confirm the
spot-interference interpretation: if spots are significant, the phase
offset is likely to change with time as the spots change. Infra-red RV
observations, which are less sensitive to spot-induced noise, would
also provide valuable further information, perhaps allowing for an
unambiguous determination of the companion mass.

Given the young age of the system, the planet candidate is likely hot,
and so a secondary transit may also be detectable with more precise
photometry, allowing constraints to be placed on the companion
temperature. Finally, simultaneous multi-band photometry and
spectroscopy (particularly in the infra-red, where spot-effects should
be lessened) could further help disentangle the signatures of star
spots and companion. This would also help to establish the cause of
the apparent overall change in transit shape between the two years'
observing runs, which could result either from changes in the spot
distribution or possibly, given the exceptionally short orbital
timescale, changes in the orbital geometry.

If our interpretation of the data is correct, the putative planet's
youth and its unique proximity to its host star will make it a
valuable object for helping inform our understanding of exoplanet
formation. Its young age could have important
implications for the mechanism by which it formed: conventional
core-accretion formation models \citep{Pollack1996} occur on
timescales of $\sim1$--$10\Myr$, comparable with or longer than the
age of the system; the much more rapid gravitational-instability
mechanism \citep{Boss1997,Boss2000} occurs on timescales orders of
magnitudes shorter, and may thus perhaps be favored \citep[see][for a
brief overview and comparison of the two mechanisms]{Baraffe2010}. The
companion's inflated atmosphere appears indicative of the ``hot start''
models of \citet{Marley2007} and \citet{Spiegel2012}, which are
associated with gravitational instability. At the same time, we cannot
rule out that formation may yet be incomplete given the young age of
the system. Neither can we rule out that the companion may be on the
verge of evaporation and may not survive the lifetime of the
star. Indeed, very few large planets are known to orbit small stars,
and this object may well be transient. Much ambiguity remains at these
young ages, and more observation and analysis of the system is
needed. If the planetary nature of the proposed companion is
confirmed, the system could provide a wealth of new information for
constraining dynamical and atmospheric evolution models for
exoplanets. We therefore propose PTFO 8-8695 as an object worthy of
further careful study by the community.

%% file: acknowledgments.tex
The authors acknowledge Tim Morton and Dimitri Veras for their helpful
input.  S.B.C.~wishes to acknowledge generous support from Gary and
Cynthia Bengier, the Richard and Rhoda Goldman Fund, National
Aeronautics and Space Administration (NASA)/\textit{Swift} grant
NNX10AI21G, NASA/\textit{Fermi} grant NNX1OA057G, and National Science
Foundation (NSF) grant AST--0908886. S.M. acknowledges support from
NSF awards AST-1006676, AST-1126413, PSARC, and the NASA Astrobiology
Institute.

Observations obtained with the Samuel Oschin Telescope at the Palomar
Observatory as part of the Palomar Transient Factory project, a
scientific collaboration between the California Institute of
Technology, Columbia University, Las Cumbres Observatory, the Lawrence
Berkeley National Laboratory, the National Energy Research Scientific
Computing Center, the University of Oxford, and the Weizmann Institute
of Science.

The Hobby-Eberly Telescope (HET) is a joint project of the University
of Texas at Austin, the Pennsylvania State University, Stanford
University, Ludwig-Maximilians-Universit\"{a}t M\"{u}nchen, and
Georg-August-Universit\"{a}t G\"{o}ttingen. The HET is named in honor
of its principal benefactors, William P. Hobby and Robert E. Eberly.

Some of the data presented herein were obtained at the W.M. Keck
Observatory, which is operated as a scientific partnership among the
California Institute of Technology, the University of California and
the National Aeronautics and Space Administration. The Observatory was
made possible by the generous financial support of the W.M. Keck
Foundation. The authors wish to recognize and acknowledge the very
significant cultural role and reverence that the summit of Mauna Kea
has always had within the indigenous Hawaiian community.  We are most
fortunate to have the opportunity to conduct observations from this
mountain.

The Byrne Observatory at Sedgwick (BOS) is operated by the Las Cumbres Observatory Global Telescope Network and is located at the Sedgwick Reserve, a part of the University of California Natural Reserve System.

This research has made use of the NASA Exoplanet Archive, which is operated by the California Institute of Technology, under contract with the National Aeronautics and Space Administration under the Exoplanet Exploration Program.

This publication makes use of data products from the Two Micron All
Sky Survey, which is a joint project of the University of
Massachusetts and the Infrared Processing and Analysis
Center/California Institute of Technology, funded by the National
Aeronautics and Space Administration and the National Science
Foundation.

This research has made use of the VizieR catalogue access tool, CDS,
Strasbourg, France; and of the software package \textit{Uncertainties:
a Python package for calculations with uncertainties}, Eric
O. Lebigot.\footnote{\url{http://packages.python.org/uncertainties}}

This work was partially supported by funding from the Center for
Exoplanets and Habitable Worlds. The Center for Exoplanets and
Habitable Worlds is supported by the Pennsylvania State University,
the Eberly College of Science, and the Pennsylvania Space Grant
Consortium.

Support for this work was provided by an award issued by JPL/Caltech.

{\it Facilities:} \facility{PO:1.2m (PTF)}, \facility{HET (HRS)},
\facility{Keck:I (HIRES)}, \facility{Keck:II (NIRC2)}, \facility{LCOGT
(BOS)}, \facility{FTN}, \facility{FTS}, \facility{Mayall},
\facility{Hale}